\documentclass[prc,aps,twocolumn,showpacs]{revtex4}
\pdfoutput=1 
\usepackage{amsmath}
\usepackage{amsfonts}
\usepackage{graphicx}
\usepackage{dcolumn}
\usepackage{hyperref}

\begin{document}


\newenvironment{ibox}[1]%
{\vskip 1.0em
\framebox[\columnwidth][r]{%
\begin{minipage}[c]{\columnwidth}%
\vspace{-1.0em}%
#1%
\end{minipage}}}
{\vskip 1.0em}

\newcommand{\iboxed}[1]{%
\vskip 1.0em
\framebox[\columnwidth][r]{%
\begin{minipage}[c]{\columnwidth}%
\vspace{-1.0em}
#1%
\end{minipage}}
\vskip 1.0em}

\newcommand{\fitbox}[2]{%
\vskip 1.0em
\begin{flushright}
\framebox[{#1}][r]{%
\begin{minipage}[c]{\columnwidth}%
\vspace{-1.0em}
#2%
\end{minipage}}
\end{flushright}
\vskip 1.0em}

\newcommand{\iboxeds}[1]{%
\vskip 1.0em
\begin{equation}
\fbox{%
\begin{minipage}[c]{1mm}%
\vspace{-1.0em}
#1%
\end{minipage}}
\end{equation}
\vskip 1.0em}

\def\Xint#1{\mathchoice
   {\XXint\displaystyle\textstyle{#1}}%
   {\XXint\textstyle\scriptstyle{#1}}%
   {\XXint\scriptstyle\scriptscriptstyle{#1}}%
   {\XXint\scriptscriptstyle\scriptscriptstyle{#1}}%
   \!\int}
\def\XXint#1#2#3{{\setbox0=\hbox{$#1{#2#3}{\int}$}
     \vcenter{\hbox{$#2#3$}}\kern-.5\wd0}}
\def\ddashint{\Xint=}
\def\dashint{\Xint-}

\newcommand{\alps}{\ensuremath{\alpha_s}}
\newcommand{\qbar}{\bar{q}}
\newcommand{\ubar}{\bar{u}}
\newcommand{\dbar}{\bar{d}}
\newcommand{\sbar}{\bar{s}}
\newcommand{\beq}{\begin{equation}}
\newcommand{\eeq}{\end{equation}}
\newcommand{\beqa}{\begin{eqnarray}}
\newcommand{\eeqa}{\end{eqnarray}}
\newcommand{\gs}{g_{\pi NN}}
\newcommand{\gw}{f_\pi}
\newcommand{\mq}{m_Q}
\newcommand{\mn}{m_N}
\newcommand{\mpi}{m_\pi}
\newcommand{\mrho}{m_\rho}
\newcommand{\momg}{m_\omega}
\newcommand{\bb}{\langle}
\newcommand{\kb}{\rangle}
\newcommand{\xvec}{\mathbf{x}}
\newcommand{\st}{\ensuremath{\sqrt{\sigma}}}
\newcommand{\Bvec}{\mathbf{B}}
\newcommand{\rvec}{\mathbf{r}}
\newcommand{\kvec}{\mathbf{k}}
\newcommand{\bvec}[1]{\ensuremath{\mathbf{#1}}}
\newcommand{\bra}[1]{\ensuremath{\bb#1|}}
\newcommand{\ket}[1]{\ensuremath{|#1\kb}}
\newcommand{\gft}{\ensuremath{\gamma_{FT}}}
\newcommand{\gfv}{\ensuremath{\gamma_5}}
\newcommand{\bfalp}{\ensuremath{\bm{\alpha}}}
\newcommand{\bfbeta}{\ensuremath{\bm{\beta}}}
\newcommand{\bfeps}{\ensuremath{\bm{\epsilon}}}
\newcommand{\lag}{{\lambda_\gamma}}
\newcommand{\lao}{{\lambda_\omega}}
\newcommand{\lN}{\lambda_N}
\newcommand{\lM}{\lambda_M}
\newcommand{\lB}{\lambda_B}
\newcommand{\epslag}{\ensuremath{\bm{\epsilon}_{\lag}}}
\newcommand{\bfept}{\ensuremath{\tilde{\bm{\epsilon}}}}
\newcommand{\bfgam}{\ensuremath{\bm{\gamma}}}
\newcommand{\bfnab}{\ensuremath{\bm{\nabla}}}
\newcommand{\bflambda}{\ensuremath{\bm{\lambda}}}
\newcommand{\bfmu}{\ensuremath{\bm{\mu}}}
\newcommand{\bfphi}{\ensuremath{\bm{\phi}}}
\newcommand{\bfvphi}{\ensuremath{\bm{\varphi}}}
\newcommand{\bfpi}{\ensuremath{\bm{\pi}}}
\newcommand{\bfsig}{\ensuremath{\bm{\sigma}}}
\newcommand{\bftau}{\ensuremath{\bm{\tau}}}
\newcommand{\bfrho}{\ensuremath{\bm{\rho}}}
\newcommand{\bfth}{\ensuremath{\bm{\theta}}}
\newcommand{\bfchi}{\ensuremath{\bm{\chi}}}
\newcommand{\bfxi}{\ensuremath{\bm{\xi}}}
\newcommand{\bfR}{\ensuremath{\bvec{R}}}
\newcommand{\bfP}{\ensuremath{\bvec{P}}}
\newcommand{\Rcm}{\ensuremath{\bvec{R}_{CM}}}
\newcommand{\spup}{\uparrow}
\newcommand{\spd}{\downarrow}
\newcommand{\up}{\uparrow}
\newcommand{\dn}{\downarrow}
\newcommand{\hbarom}{\frac{\hbar^2}{m_Q}}
\newcommand{\half}{\ensuremath{\frac{1}{2}}}
\newcommand{\thalf}{\ensuremath{\frac{3}{2}}}
\newcommand{\fhalf}{\ensuremath{\frac{5}{2}}}
\newcommand{\shalf}{\ensuremath{{\tfrac{1}{2}}}}
\newcommand{\sqtr}{\ensuremath{{\tfrac{1}{4}}}}
\newcommand{\sphalf}{\ensuremath{\genfrac{}{}{0pt}{1}{+}{}\!\tfrac{1}{2}}}
\newcommand{\smhalf}{\ensuremath{\genfrac{}{}{0pt}{1}{-}{}\!\tfrac{1}{2}}}
\newcommand{\sthalf}{\ensuremath{{\tfrac{3}{2}}}}
\newcommand{\spthalf}{\ensuremath{{\tfrac{+3}{2}}}}
\newcommand{\smthalf}{\ensuremath{{\tfrac{-3}{2}}}}
\newcommand{\sfhalf}{{\tfrac{5}{2}}}
\newcommand{\third}{{\frac{1}{3}}}
\newcommand{\tthird}{{\frac{2}{3}}}
\newcommand{\sthird}{{\tfrac{1}{3}}}
\newcommand{\stthird}{{\tfrac{2}{3}}}
\newcommand{\vnn}{\ensuremath{\hat{v}_{NN}}}
\newcommand{\vij}{\ensuremath{\hat{v}_{ij}}}
\newcommand{\vik}{\ensuremath{\hat{v}_{ik}}}
\newcommand{\argonne}{\ensuremath{v_{18}}}
\newcommand{\lqcd}{\ensuremath{\mathcal{L}_{QCD}}}
\newcommand{\lqed}{\ensuremath{\mathscr{L}_{QED}}}
\newcommand{\lgf}{\ensuremath{\mathcal{L}_g}}
\newcommand{\lqm}{\ensuremath{\mathcal{L}_q}}
\newcommand{\lqg}{\ensuremath{\mathcal{L}_{qg}}}
\newcommand{\nn}{\ensuremath{N\!N}}
\newcommand{\nnn}{\ensuremath{N\!N\!N}}
\newcommand{\qq}{\ensuremath{qq}}
\newcommand{\qqq}{\ensuremath{qqq}}
\newcommand{\qqb}{\ensuremath{q\bar{q}}}
\newcommand{\hpnd}{\ensuremath{H_{\pi N\Delta}}}
\newcommand{\hpqq}{\ensuremath{H_{\pi qq}}}
\newcommand{\hpqqa}{\ensuremath{H^{(a)}_{\pi qq}}}
\newcommand{\hpqqe}{\ensuremath{H^{(e)}_{\pi qq}}}
\newcommand{\hint}{\ensuremath{H_{\rm int}}}
\newcommand{\fpnn}{\ensuremath{f_{\pi\! N\!N}}}
\newcommand{\fenn}{\ensuremath{f_{\eta\! N\!N}}}
\newcommand{\gsnn}{\ensuremath{g_{\sigma\! N\!N}}}
\newcommand{\gpnn}{\ensuremath{g_{\pi\! N\!N}}}
\newcommand{\fpnd}{\ensuremath{f_{\pi\! N\!\Delta}}}
\newcommand{\grpg}{\ensuremath{g_{\rho\pi\gamma}}}
\newcommand{\gopg}{\ensuremath{g_{\omega\pi\gamma}}}
\newcommand{\fmqq}{\ensuremath{f_{M\! qq}}}
\newcommand{\gmqq}{\ensuremath{g_{M\! qq}}}
\newcommand{\fpqq}{\ensuremath{f_{\pi qq}}}
\newcommand{\gpqq}{\ensuremath{g_{\pi qq}}}
\newcommand{\feqq}{\ensuremath{f_{\eta qq}}}
\newcommand{\gonn}{\ensuremath{g_{\omega N\!N}}}
\newcommand{\gonna}{\ensuremath{g^t_{\omega N\!N}}}
\newcommand{\grnn}{\ensuremath{g_{\rho N\!N}}}
\newcommand{\gopr}{\ensuremath{g_{\omega\pi\rho}}}
\newcommand{\grnp}{\ensuremath{g_{\rho N\!\pi}}}
\newcommand{\grpp}{\ensuremath{g_{\rho\pi\pi}}}
\newcommand{\Lpnn}{\ensuremath{\Lambda_{\pi\! N\! N}}}
\newcommand{\Lonn}{\ensuremath{\Lambda_{\omega N\! N}}}
\newcommand{\Lonna}{\ensuremath{\Lambda^t_{\omega N\! N}}}
\newcommand{\Lrnn}{\ensuremath{\Lambda_{\rho N\! N}}}
\newcommand{\Lopr}{\ensuremath{\Lambda_{\omega\pi\rho}}}
\newcommand{\Lrpp}{\ensuremath{\Lambda_{\rho\pi\pi}}}
\newcommand{\getaqq}{\ensuremath{g_{\eta qq}}}
\newcommand{\fsqq}{\ensuremath{f_{\sigma qq}}}
\newcommand{\gsqq}{\ensuremath{g_{\sigma qq}}}
\newcommand{\piqq}{\ensuremath{{\pi\! qq}}}
\newcommand{\ylm}{\ensuremath{Y_\ell^m}}
\newcommand{\ylmc}{\ensuremath{Y_\ell^{m*}}}
\newcommand{\ebh}[1]{\hat{\bvec{e}}_{#1}}
\newcommand{\kbh}{\hat{\bvec{k}}}
\newcommand{\nbh}{\hat{\bvec{n}}}
\newcommand{\pvbh}{\hat{\bvec{p}}}
\newcommand{\qbh}{\hat{\bvec{q}}}
\newcommand{\Xbh}{\hat{\bvec{X}}}
\newcommand{\rbh}{\hat{\bvec{r}}}
\newcommand{\xbh}{\hat{\bvec{x}}}
\newcommand{\ybh}{\hat{\bvec{y}}}
\newcommand{\zbh}{\hat{\bvec{z}}}
\newcommand{\betabh}{\hat{\bfbeta}}
\newcommand{\tbh}{\hat{\bfth}}
\newcommand{\pbh}{\hat{\bfvphi}}
\newcommand{\dt}{\Delta\tau}
\newcommand{\kmag}{|\bvec{k}|}
\newcommand{\pmag}{|\bvec{p}|}
\newcommand{\qmag}{|\bvec{q}|}
\newcommand{\oas}{\ensuremath{\mathcal{O}(\alpha_s)}}
\newcommand{\vtxb}{\ensuremath{\Lambda_\mu(p',p)}}
\newcommand{\vtxp}{\ensuremath{\Lambda^\mu(p',p)}}
\newcommand{\pwqp}{e^{i\bvec{q}\cdot\bvec{r}}}
\newcommand{\pwqm}{e^{-i\bvec{q}\cdot\bvec{r}}}
\newcommand{\gsa}[1]{\ensuremath{\bb#1\kb_0}}
\newcommand{\oer}[1]{\mathcal{O}\left(\frac{1}{\qmag^{#1}}\right)}
\newcommand{\nub}[1]{\overline{\nu^{#1}}}
\newcommand{\epf}{E_\bvec{p}}
\newcommand{\epfp}{E_{\bvec{p}'}}
\newcommand{\eka}{E_{\alpha\kappa}}
\newcommand{\ekaq}{(E_{\alpha\kappa})^2}
\newcommand{\ekap}{E_{\alpha'\kappa}}
\newcommand{\ekpa}{E+{\alpha\kappa_+}}
\newcommand{\ekma}{E_{\alpha\kappa_-}}
\newcommand{\ekp}{E_{\kappa_+}}
\newcommand{\ekm}{E_{\kappa_-}}
\newcommand{\ekpap}{E_{\alpha'\kappa_+}}
\newcommand{\ekmap}{E_{\alpha'\kappa_-}}
\newcommand{\yjm}[1]{\mathcal{Y}_{jm}^{#1}}
\newcommand{\ysa}[3]{\mathcal{Y}_{#1,#2}^{#3}}
\newcommand{\yss}[2]{\mathcal{Y}_{#1}^{#2}}
\newcommand{\Dj}{\ensuremath{\mathscr{D}}}
\newcommand{\ysc}{\tilde{y}}
\newcommand{\enm}{\varepsilon_{NM}}
\newcommand{\Scg}[6]
	{\ensuremath{S^{#1}_{#4}\:\vphantom{S}^{#2}_{#5}
 	 \:\vphantom{S}^{#3}_{#6}\,}}
\newcommand{\Kmat}[6]
	{\ensuremath{K\left[\begin{array}{ccc} 
	#1 & #2 & #3 \\ #4 & #5 & #6\end{array}\right]}}
\newcommand{\irt}{\ensuremath{\frac{1}{\sqrt{2}}}}
\newcommand{\sirt}{\ensuremath{\tfrac{1}{\sqrt{2}}}}
\newcommand{\irth}{\ensuremath{\frac{1}{\sqrt{3}}}}
\newcommand{\sirth}{\ensuremath{\tfrac{1}{\sqrt{3}}}}
\newcommand{\irs}{\ensuremath{\frac{1}{\sqrt{6}}}}
\newcommand{\sirs}{\ensuremath{\tfrac{1}{\sqrt{6}}}}
\newcommand{\tors}{\ensuremath{\frac{2}{\sqrt{6}}}}
\newcommand{\stors}{\ensuremath{\tfrac{2}{\sqrt{6}}}}
\newcommand{\rtoth}{\ensuremath{\sqrt{\frac{2}{3}}}}
\newcommand{\rthot}{\ensuremath{\frac{\sqrt{3}}{2}}}
\newcommand{\ithrt}{\ensuremath{\frac{1}{3\sqrt{2}}}}
\newcommand{\Tg}{\ensuremath{\mathsf{T}}}
\newcommand{\irrep}[1]{\ensuremath{\mathbf{#1}}}
\newcommand{\cirrep}[1]{\ensuremath{\overline{\mathbf{#1}}}}
\newcommand{\Fij}{\ensuremath{\hat{F}_{ij}}}
\newcommand{\Fqij}{\ensuremath{\hat{F}^{(qq)}_{ij}}}
\newcommand{\Fsij}{\ensuremath{\hat{F}^{(qs)}_{ij}}}
\newcommand{\Opij}{\mathcal{O}^p_{ij}}
\newcommand{\fpij}{f_p(r_{ij})}
\newcommand{\titj}{\bftau_i\cdot\bftau_j}
\newcommand{\sisj}{\bfsig_i\cdot\bfsig_j}
\newcommand{\Sij}{S_{ij}}
\newcommand{\LS}{\bvec{L}_{ij}\cdot\bvec{S}_{ij}}
\newcommand{\TT}{\Tg_i\cdot\Tg_j}
\newcommand{\chet}{\ensuremath{\chi ET}}
\newcommand{\chpt}{\ensuremath{\chi PT}}
\newcommand{\chsy}{\ensuremath{\chi\mbox{symm}}}
\newcommand{\lchi}{\ensuremath{\Lambda_\chi}}
\newcommand{\lcon}{\ensuremath{\Lambda_{QCD}}}
\newcommand{\dcpsi}{\ensuremath{\bar{\psi}}}
\newcommand{\dc}[1]{\ensuremath{\overline{#1}}}
\newcommand{\dcpsip}{\ensuremath{\bar{\psi}^{(+)}}}
\newcommand{\psip}{\ensuremath{{\psi}^{(+)}}}
\newcommand{\dcpsim}{\ensuremath{\bar{\psi}^{(-)}}}
\newcommand{\psim}{\ensuremath{{\psi}^{(-)}}}
\newcommand{\llo}{\ensuremath{\mathcal{L}^{(0)}_{\chet}}}
\newcommand{\lchet}{\ensuremath{\mathcal{L}_{\chi}}}
\newcommand{\hchet}{\ensuremath{\mathcal{H}_{\chi}}}
\newcommand{\Hd}{\ensuremath{\mathcal{H}}}
\newcommand{\Dmu}{\ensuremath{\mathcal{D}_\mu}}
\newcommand{\Dsl}{\ensuremath{\slashed{\mathcal{D}}}}
\newcommand{\comm}[2]{\ensuremath{\left[#1,#2\right]}}
\newcommand{\acomm}[2]{\ensuremath{\left\{#1,#2\right\}}}
\newcommand{\ev}[1]{\ensuremath{\bb\hat{#1}\kb}}
\newcommand{\evt}[1]{\ensuremath{\bb{#1}(\tau)\kb}}
\newcommand{\evm}[1]{\ensuremath{\bb{#1}\kb_M}}
\newcommand{\evv}[1]{\ensuremath{\bb{#1}\kb_V}}
\newcommand{\ovl}[2]{\ensuremath{\bb{#1}|{#2}\kb}}
\newcommand{\pd}{\partial}
\newcommand{\pnpd}[2]{\frac{\partial{#1}}{\partial{#2}}}
\newcommand{\pppd}[1]{\frac{\partial{\hphantom{#1}}}{\partial{#1}}}
\newcommand{\plmu}{\partial_\mu}
\newcommand{\plnu}{\partial_\nu}
\newcommand{\pumu}{\partial^\mu}
\newcommand{\punu}{\partial^\nu}
\newcommand{\mcdf}{\delta^{(4)}(p_f-p_i-q)}
\newcommand{\ecdf}{\delta(E_f-E_i-\nu)}
\newcommand{\tr}{\mbox{Tr }}
\newcommand{\lxr}{\ensuremath{SU(2)_L\times SU(2)_R}}
\newcommand{\gV}[2]{\ensuremath{(\gamma^{-1})^{#1}_{\hphantom{#1}{#2}}}}
\newcommand{\gVd}[2]{\ensuremath{\gamma^{#1}_{\hphantom{#1}{#2}}}}
\newcommand{\LpV}[1]{\ensuremath{\Lambda^{#1}V}}
\newcommand{\hatH}{\ensuremath{\hat{H}}}
\newcommand{\hath}{\ensuremath{\hat{h}}}
\newcommand{\eht}{\ensuremath{e^{-\tau\hat{H}}}}
\newcommand{\ehdt}{\ensuremath{e^{-\Delta\tau\hat{H}}}}
\newcommand{\ehtm}{\ensuremath{e^{-\tau(\hat{H}-E_V)}}}
\newcommand{\ehdtm}{\ensuremath{e^{-\Delta\tau(\hat{H}-E_V)}}}
\newcommand{\Oop}{\ensuremath{\mathcal{O}}}
\newcommand{\Gop}{\ensuremath{\hat{\mathcal{G}}}}
\newcommand{\SU}[1]{\ensuremath{SU({#1})}}
\newcommand{\U}[1]{\ensuremath{U({#1})}}
\newcommand{\proj}[1]{\ensuremath{\ket{#1}\bra{#1}}}
\newcommand{\su}[1]{\ensuremath{\mathfrak{su}({#1})}}
\newcommand{\ip}[2]{\ensuremath{\bvec{#1}\cdot\bvec{#2}}}
\newcommand{\norm}[1]{\ensuremath{\left| #1\right|^2}}
\newcommand{\rnorm}[1]{\ensuremath{\lvert #1\rvert}}
\newcommand{\pid}{\left(\begin{array}{cc} 1 & 0 \\ 0 & 1\end{array}\right)}
\newcommand{\psx}{\left(\begin{array}{cc} 0 & 1 \\ 1 & 0\end{array}\right)}
\newcommand{\psy}{\left(\begin{array}{cc} 0 & -i \\ i & 0\end{array}\right)}
\newcommand{\psz}{\left(\begin{array}{cc} 1 & 0 \\ 0 & -1\end{array}\right)}
\newcommand{\ua}{\uparrow}
\newcommand{\da}{\downarrow}
\newcommand{\deln}{\delta_{i_1 i_2\ldots i_n}}
\newcommand{\GabRR}{G_{\alpha\beta}(\bfR,\bfR')}
\newcommand{\GRR}{G(\bfR,\bfR')}
\newcommand{\GfRR}{G_0(\bfR,\bfR')}
\newcommand{\GRiR}{G(\bfR_i,\bfR_{i-1})}
\newcommand{\GRRs}[2]{G(\bfR_{#1},\bfR_{#2})}
\newcommand{\Gdgn}{\Gamma_{\Delta,\gamma N}}
\newcommand{\Gdgnb}{\overline\Gamma_{\Delta,\gamma N}}
\newcommand{\GJT}{\Gamma_{LS}^{JT}(k)}
\newcommand{\GJTa}[2]{\Gamma^{#1}_{#2}}
\newcommand{\GtwJTa}[2]{\tilde{\Gamma}_{#1}^{#2}}
\newcommand{\Gtw}{\tilde{\Gamma}}
\newcommand{\Gbar}{\overline{\Gamma}}
\newcommand{\Gtil}{\tilde{\Gamma}}
\newcommand{\Gpndb}{\overline{\Gamma}_{\pi N,\Delta}}
\newcommand{\GbNgn}{{\overline{\Gamma}}_{N^*,\gamma N}}
\newcommand{\GNgn}{\Gamma_{N^*,\gamma N}}
\newcommand{\GbNmb}{{\overline{\Gamma}}_{N^*,MB}}
\newcommand{\Lg}[2]{\ensuremath{L^{#1}_{\hphantom{#1}{#2}}}}
\newcommand{\psik}{\ensuremath{\left(\begin{matrix}\psi_1 \\ \psi_2\end{matrix}\right)}}
\newcommand{\psib}{\ensuremath{\left(\begin{matrix}\psi^*_1&\psi^*_2\end{matrix}\right)}}
\newcommand{\Gf}{\ensuremath{\frac{1}{E-H_0}}}
\newcommand{\Gv}{\ensuremath{\frac{1}{E-H_0-\vnres}}}
\newcommand{\Gx}{\ensuremath{\frac{1}{E-H_0-V}}}
\newcommand{\Gex}{\ensuremath{\mathcal{G}}}
\newcommand{\Gfpm}{\ensuremath{\frac{1}{E-H_0\pm i\epsilon}}}
\newcommand{\vres}{v_R}
\newcommand{\vnres}{v}
\newcommand{\tpz}{\ensuremath{^3P_0}}
\newcommand{\tres}{t_R}
\newcommand{\tsr}{t^R}
\newcommand{\tsnr}{t^{NR}}
\newcommand{\trest}{\tilde{t}_R}
\newcommand{\tnres}{t}
\newcommand{\Pt}{P_{12}}
\newcommand{\Sz}{\ket{S_0}}
\newcommand{\Sa}{\ket{S^{(-1)}_1}}
\newcommand{\Sb}{\ket{S^{(0)}_1}}
\newcommand{\Sc}{\ket{S^{(+1)}_1}}
\newcommand{\sbasis}{\ket{s_1 s_2; m_1 m_2}}
\newcommand{\Sbasis}{\ket{s_1 s_2; S M}}
\newcommand{\sket}[2]{\ket{{#1}\,{#2}}}
\newcommand{\sbra}[2]{\bra{{#1}\,{#2}}}
\newcommand{\psmket}{\ket{\bvec{p};s\,m}}
\newcommand{\cket}{\ket{\bvec{p};s_1 s_2\,m_1 m_2}}
\newcommand{\hket}{\ket{\bvec{p};s_1 s_2\,\lambda_1\lambda_2}}
\newcommand{\hkets}{\ket{s\,\lambda}}
\newcommand{\phkets}{\ket{\bvec{p};s\,\lambda}}
\newcommand{\klsjm}{\ket{p;\ell s; j m}}
\newcommand{\pq}{\bvec{p}_q}
\newcommand{\pqb}{\bvec{p}_{\qbar}}
\newcommand{\mps}[1]{\frac{d^3{#1}}{(2\pi)^{3/2}}}
\newcommand{\mpsf}[1]{\frac{d^3{#1}}{(2\pi)^{3}}}
\newcommand{\du}[1]{u_{\bvec{#1},s}}
\newcommand{\dv}[1]{v_{\bvec{#1},s}}
\newcommand{\cdu}[1]{\overline{u}_{\bvec{#1},s}}
\newcommand{\cdv}[1]{\overline{v}_{\bvec{#1},s}}
\newcommand{\dus}[2]{u_{\bvec{#1},{#2}}}
\newcommand{\dvs}[2]{v_{\bvec{#1},{#2}}}
\newcommand{\cdus}[2]{\overline{u}_{\bvec{#1},{#2}}}
\newcommand{\cdvs}[2]{\overline{v}_{\bvec{#1},{#2}}}
\newcommand{\bop}[1]{b_{\bvec{#1},s}}
\newcommand{\dop}[1]{d_{\bvec{#1},s}}
\newcommand{\bops}[2]{b_{\bvec{#1},{#2}}}
\newcommand{\dops}[2]{d_{\bvec{#1},{#2}}}
\newcommand{\mev}{\mbox{ MeV}}
\newcommand{\gev}{\mbox{ GeV}}
\newcommand{\fmi}{\mbox{ fm}}
\newcommand{\M}{\mathcal{M}}
\newcommand{\Smat}{\mathcal{S}}
\newcommand{\JLSTh}{JLST\lambda}
\newcommand{\Tpg}{T_{\pi N,\gamma N}}
\newcommand{\tpg}{t_{\pi N,\gamma N}}
\newcommand{\vmbmb}{\ensuremath{v_{M'B',MB}}}
\newcommand{\tmbgn}{\ensuremath{t_{MB,\gamma N}}}
\newcommand{\Tonon}{\ensuremath{T_{\omega N,\omega N}}}
\newcommand{\tonon}{\ensuremath{t_{\omega N,\omega N}}}
\newcommand{\tronon}{\ensuremath{t^R_{\omega N,\omega N}}}
\newcommand{\Tonpn}{\ensuremath{T_{\omega N,\pi N}}}
\newcommand{\tonpn}{\ensuremath{t_{\omega N,\pi N}}}
\newcommand{\tronpn}{\ensuremath{t^R_{\omega N,\pi N}}}
\newcommand{\Tongn}{\ensuremath{T_{\omega N,\gamma N}}}
\newcommand{\tongn}{\ensuremath{t_{\omega N,\gamma N}}}
\newcommand{\trongn}{\ensuremath{t^R_{\omega N,\gamma N}}}
\newcommand{\vmbgn}{\ensuremath{v_{MB,\gamma N}}}
\newcommand{\vpngn}{\ensuremath{v_{\pi N,\gamma N}}}
\newcommand{\vongn}{\ensuremath{v_{\omega N,\gamma N}}}
\newcommand{\vonpn}{\ensuremath{v_{\omega N,\pi N}}}
\newcommand{\vpnpn}{\ensuremath{v_{\pi N,\pi N}}}
\newcommand{\vonon}{\ensuremath{v_{\omega N,\omega N}}}
\newcommand{\vrngn}{\ensuremath{v_{\rho N,\gamma N}}}
\newcommand{\tjtmbmb}{\ensuremath{t^{JT}_{M'B',MB}}}
\newcommand{\tjlsmngn}{\ensuremath{t^{JT}_{L'S'M'N',\lag\lN T_{N,z}}}}
\newcommand{\tjlsmbgn}{\ensuremath{t^{JT}_{LSMB,\lag \lN T_{N,z}}}}
\newcommand{\vjlsmngn}{\ensuremath{v^{JT}_{L'S'M'N',\lag \lN T_{N,z}}}}
\newcommand{\vjlsmbgn}{\ensuremath{v^{JT}_{LSMB,\lag \lN T_{N,z}}}}
\newcommand{\tjlsmnmb}{\ensuremath{t^{JT}_{L'S'M'N',LSMB}}}
\newcommand{\Tjlsmbmb}{\ensuremath{T^{JT}_{LSMB,L'S'M'B'}}}
\newcommand{\tjlsmbmb}{\ensuremath{t^{JT}_{LSMB,L'S'M'B'}}}
\newcommand{\tjlsmnpn}{\ensuremath{t^{JT}_{L'S'M'N',\ell \pi N}}}
\newcommand{\tjlsmbpn}{\ensuremath{t^{JT}_{LSMB,\ell \pi N}}}
\newcommand{\vjlsmnpn}{\ensuremath{v^{JT}_{L'S'M'N',\ell \pi N}}}
\newcommand{\vjlsmnmb}{\ensuremath{v^{JT}_{L'S'M'N',LSMB}}}
\newcommand{\vjlsmbpn}{\ensuremath{v^{JT}_{LSMB,\ell \pi N}}}
\newcommand{\Tjlsmngn}{\ensuremath{t^{R,JT}_{L'S'M'N',\lag\lN T_{N,z}}}}
\newcommand{\Tjlsmbgn}{\ensuremath{t^{R,JT}_{LSMB,\lag \lN T_{N,z}}}}
\newcommand{\Tfjlsmbgn}{\ensuremath{T^{JT}_{LSMB,\lag \lN T_{N,z}}}}
\newcommand{\Tjlsmnmb}{\ensuremath{t^{R,JT}_{L'S'M'N',LSMB}}}
\newcommand{\Tjlsmnpn}{\ensuremath{t^{R,JT}_{L'S'M'N',\ell \pi N}}}
\newcommand{\Tjlsmbpn}{\ensuremath{t^{R,JT}_{LSMB,\ell \pi N}}}
\newcommand{\Gbjlsi}{\ensuremath{{\Gamma}^{JT}_{LSMB,N^*_i}}}
\newcommand{\Gbjlspi}{\ensuremath{{\Gamma}^{JT}_{L'S'M'B',N^*_i}}}
\newcommand{\Gjlsi}{\ensuremath{\overline{\Gamma}^{JT}_{LSMB,N^*_i}}}
\newcommand{\Gijls}{\ensuremath{\overline{\Gamma}^{JT}_{N^*_i,LSMB}}}
\newcommand{\Gbijls}{\ensuremath{{\Gamma}^{JT}_{N^*_i,LSMB}}}
\newcommand{\Gjpn}{\ensuremath{\overline{\Gamma}^{JT}_{N^*_j,\ell\pn}}}
\newcommand{\Gign}{\ensuremath{\overline{\Gamma}^{JT}_{N^*_i,\lag\lN T_{N,z}}}}
\newcommand{\Gbign}{\ensuremath{{\Gamma}^{JT}_{N^*_i,\lag\lN T_{N,z}}}}
\newcommand{\Gjlsj}{\ensuremath{\overline{\Gamma}^{JT}_{LSMB,N^*_j}}}
\newcommand{\Gjem}{\ensuremath{\overline{\Gamma}^{JT}_{N^*_j,\lag\lN T_{N,z}}}}
\newcommand{\Ljtlsmbn}{\ensuremath{\Lambda^{JT}_{N^*LSMB}}}
\newcommand{\Drij}{\ensuremath{\mathcal{D}^{-1}_{ij}}}
\newcommand{\Mbres}{\ensuremath{M^{(0)}_{N^*}}}
\newcommand{\Cjtnlsmb}{\ensuremath{C^{JT}_{N^*LSMB}}}
\newcommand{\Ljtnlsmb}{\ensuremath{\Lambda^{JT}_{N^*LSMB}}}
\newcommand{\knstar}{\ensuremath{k_{N^*}}}
\newcommand{\vonen}{\ensuremath{v_{\omega N,\eta N}}}
\newcommand{\vonpd}{\ensuremath{v_{\omega N,\pi\Delta}}}
\newcommand{\vonsn}{\ensuremath{v_{\omega N,\sigma N}}}
\newcommand{\vonrn}{\ensuremath{v_{\omega N,\rho N}}}
\newcommand{\gnon}{\ensuremath{\gamma N\to \omega N}}
\newcommand{\gnpn}{\ensuremath{\gamma N\to \pi N}}
\newcommand{\gpop}{\ensuremath{\gamma p\to \omega p}}
\newcommand{\gppzp}{\ensuremath{\gamma p\to \pi^0 p}}
\newcommand{\gpppn}{\ensuremath{\gamma p\to \pi^+ n}}
\newcommand{\pnon}{\ensuremath{\pi N\to \omega N}}
\newcommand{\pnmb}{\ensuremath{\pi N\to MB}}
\newcommand{\gnmb}{\ensuremath{\gamma N\to M\!B}}
\newcommand{\onon}{\ensuremath{\omega N\to \omega N}}
\newcommand{\pmpon}{\ensuremath{\pi^- p\to \omega n}}
\newcommand{\pnpn}{\ensuremath{\pi N\to \pi N}}
\newcommand{\Gon}{\ensuremath{G_{0,\omega N}}}
\newcommand{\Gpn}{\ensuremath{G_{0,\pi N}}}
\newcommand{\rhomb}{\ensuremath{\rho_{MB}}}
\newcommand{\rhoon}{\ensuremath{\rho_{\omega N}}}
\newcommand{\rhopn}{\ensuremath{\rho_{\pi N}}}
\newcommand{\kon}{\ensuremath{k_{\omega N}}}
\newcommand{\kpn}{\ensuremath{k_{\pi N}}}
\newcommand{\Gmb}{\ensuremath{G_{0,MB}}}
\newcommand{\Tmbgn}{\ensuremath{T_{MB,\gamma N}}}
\newcommand{\vmbpgn}{\ensuremath{v_{M'B',\gamma N}}}
\newcommand{\pntpn}{\ensuremath{\pi N\!\to\!\pi N}}
\newcommand{\pnten}{\ensuremath{\pi N\!\to\!\eta N}}
\newcommand{\pnton}{\ensuremath{\pi N\!\to\!\omega N}}
\newcommand{\epos}{\ensuremath{\slashed{\epsilon}_{\lambda_\omega}}}
\newcommand{\epo}{\ensuremath{{\epsilon}_{\lambda_\omega}}}
\newcommand{\elevi}{\ensuremath{{\epsilon}_{\alpha\beta\gamma\delta}}}
\newcommand{\eps}{\ensuremath{\epsilon}}
\newcommand{\krho}{\ensuremath{\kappa_\rho}}
\newcommand{\komg}{\ensuremath{\kappa_\omega}}
\newcommand{\komga}{\ensuremath{\kappa^t_\omega}}
\newcommand{\doh}{\ensuremath{d^{(\half)}_{\lambda'\lambda}}}
\newcommand{\dohm}{\ensuremath{d^{(\half)}_{-\lambda,-\lambda'}}}
\newcommand{\dohmo}{\ensuremath{d^{(\half)}_{\lambda',-\half}}}
\newcommand{\dohpo}{\ensuremath{d^{(\half)}_{\lambda',+\half}}}
\newcommand{\Lor}[2]{\ensuremath{\Lambda^{#1}_{\hphantom{#1}{#2}}}}
\newcommand{\ILor}[2]{\ensuremath{\Lambda_{#1}^{\hphantom{#1}{#2}}}}
\newcommand{\LorT}[2]{\ensuremath{[\Lambda^T]^{#1}_{\hphantom{#1}{#2}}}}
\newcommand{\dsdo}{{\frac{d\sigma}{d\Omega}}}
\newcommand{\dspdo}{\ensuremath{{\frac{d\sigma_\pi}{d\Omega}}}}
\newcommand{\dsgdo}{\ensuremath{{\frac{d\sigma_\gamma}{d\Omega}}}}
\newcommand{\chipd}{\ensuremath{\chi^2/N_d}}
\newcommand{\chipda}{\ensuremath{\chi^2(\alpha)/N_d}}
\newcommand{\bpop}{\ensuremath{\bvec{p}'_1}}
\newcommand{\bptp}{\ensuremath{\bvec{p}'_2}}
\newcommand{\bpip}{\ensuremath{\bvec{p}'_i}}
\newcommand{\bpo}{\ensuremath{\bvec{p}_1}}
\newcommand{\bpt}{\ensuremath{\bvec{p}_2}}
\newcommand{\bpi}{\ensuremath{\bvec{p}_i}}
\newcommand{\bqo}{\ensuremath{\bvec{q}_1}}
\newcommand{\bqt}{\ensuremath{\bvec{q}_2}}
\newcommand{\bqi}{\ensuremath{\bvec{q}_i}}
\newcommand{\bQ}{\ensuremath{\bvec{Q}}}
\newcommand{\bq}{\ensuremath{\bvec{q}}}
\newcommand{\ketq}{\ensuremath{\ket{\bqo,\bqt}}}
\newcommand{\ketqc}{\ensuremath{\ket{\bQ,\bq}}}
\newcommand{\bP}{\ensuremath{\bvec{P}}}
\newcommand{\bPp}{\ensuremath{\bvec{P}'}}
\newcommand{\bpr}{\ensuremath{\bvec{p}}}
\newcommand{\bprp}{\ensuremath{\bvec{p}'}}
\newcommand{\ketPsiq}{\ensuremath{\ket{\Psi_{\bq}^{(\pm)}}}}
\newcommand{\ketPsiqQ}{\ensuremath{\ket{\Psi_{\bQ,\bq}^{(\pm)}}}}
\newcommand{\Ld}{\ensuremath{\mathcal{L}}}
\newcommand{\ps}{\mbox{ps}}
\newcommand{\fndp}{f_{N\Delta\pi}}
\newcommand{\fndr}{f_{N\Delta\rho}}
\newcommand{\said}{{\sc said}}
\newcommand{\ret}{\ensuremath{\langle{\tt ret}\rangle}}
\newcommand{\ddf}[1]{\ensuremath{\delta^{(#1)}}}
\newcommand{\Tpp}{\ensuremath{T_{\pi\pi}}}
\newcommand{\Kpp}{\ensuremath{K_{\pi\pi}}}
\newcommand{\Tpe}{\ensuremath{T_{\pi\eta}}}
\newcommand{\Kpe}{\ensuremath{K_{\pi\eta}}}
\newcommand{\Tep}{\ensuremath{T_{\eta\pi}}}
\newcommand{\Kep}{\ensuremath{K_{\eta\pi}}}
\newcommand{\Tee}{\ensuremath{T_{\eta\eta}}}
\newcommand{\Kee}{\ensuremath{K_{\eta\eta}}}
\newcommand{\Tpig}{\ensuremath{T_{\pi\gamma}}}
\newcommand{\Kpig}{\ensuremath{K_{\pi\gamma}}}
\newcommand{\oKpig}{\ensuremath{\overline{K}_{\pi\gamma}}}
\newcommand{\tKpig}{\ensuremath{\tilde{K}_{\pi\gamma}}}
\newcommand{\Teg}{\ensuremath{T_{\eta\gamma}}}
\newcommand{\Keg}{\ensuremath{K_{\eta\gamma}}}
\newcommand{\Kab}{\ensuremath{K_{\alpha\beta}}}
\newcommand{\R}{\ensuremath{\mathbb{R}}}
\newcommand{\C}{\ensuremath{\mathbb{C}}}
\newcommand{\Ezp}{\ensuremath{E^{\pi}_{0+}}}
\newcommand{\Eze}{\ensuremath{E^{\eta}_{0+}}}
\newcommand{\Ga}{\ensuremath{\Gamma_\alpha}}
\newcommand{\Gb}{\ensuremath{\Gamma_\beta}}
\newcommand{\RH}{\ensuremath{\mathcal{R}\!\!-\!\!\mathcal{H}}}
\newcommand{\calT}{\mathcal{T}}
\newcommand{\maid}{{\sc maid}}
\newcommand{\Kbar}{\ensuremath{\overline{K}}}
\newcommand{\zbar}{\ensuremath{\overline{z}}}
\newcommand{\kbar}{\ensuremath{\overline{k}}}
\newcommand{\dom}{\ensuremath{\mathcal{D}}}
\newcommand{\domi}[1]{\ensuremath{\mathcal{D}_{#1}}}
\newcommand{\pbar}{\ensuremath{\overline{p}}}
\newcommand{\Nab}{\ensuremath{N_{\alpha\beta}}}
\newcommand{\Nee}{\ensuremath{N_{\eta\eta}}}
\newcommand{\dth}[1]{\delta^{(3)}(#1)}
\newcommand{\dfo}[1]{\delta^{(4)}(#1)}
\newcommand{\intk}{\int\!\!\frac{d^3\! k}{(2\pi)^3}}
\newcommand{\intkg}{\int\!\!{d^3\! k_\gamma}}
\newcommand{\intks}{\int\!\!{d^3\! k_\sigma}}
\newcommand{\nch}{\ensuremath{N_{\mbox{ch}}}}
\newcommand{\nc}{\ensuremath{N_{ch}}}
\newcommand{\re}{\ensuremath{\mbox{Re }\!}}
\newcommand{\im}{\ensuremath{\mbox{Im }\!}}
\newcommand{\EetaS}{\ensuremath{E^\eta_{0+}}}
\newcommand{\EpiS}{\ensuremath{E^\pi_{0+}}}

\newcommand{\gn}{\ensuremath{\gamma N}}
\newcommand{\gp}{\ensuremath{\gamma p}}
\newcommand{\geta}{\ensuremath{\gamma \eta}}
\newcommand{\pp}{\ensuremath{pp}}
\newcommand{\pn}{\ensuremath{\pi N}}
\newcommand{\phn}{\ensuremath{\pi d}}
\newcommand{\en}{\ensuremath{\eta N}}
\newcommand{\epn}{\ensuremath{\eta' N}}
\newcommand{\pD}{\ensuremath{\pi \Delta}}
\newcommand{\sn}{\ensuremath{\sigma N}}
\newcommand{\rn}{\ensuremath{\rho N}}
\newcommand{\on}{\ensuremath{\omega N}}
\newcommand{\ppn}{\ensuremath{\pi\pi N}}
\newcommand{\kn}{\ensuremath{KN}}
\newcommand{\ky}{\ensuremath{KY}}
\newcommand{\kl}{\ensuremath{K\Lambda}}
\newcommand{\ks}{\ensuremath{K\Sigma}}
\newcommand{\bn}{\ensuremath{eN}}
\newcommand{\bpn}{\ensuremath{e\pi N}}
\newcommand{\fpo}{\ensuremath{5\oplus 1}}
\newcommand{\faoe}{{\sc FA08}}
\newcommand{\fpoe}{{\sc FP08}}
\newcommand{\fsoe}{{\sc FS08}}

\newcommand{\itPFP}{\textit{Physics for Future Presidents}}
\newcommand{\itaPFP}{\textit{PFP}}

\title{Toward a unified description of hadro- and photoproduction:
$S$--wave $\pi$-- and $\eta$--photoproduction amplitudes}
\author{Mark W.\ Paris and Ron L.\ Workman}
\affiliation{
Data Analysis Center at the Center for Nuclear Studies,\\
Department of Physics\\
The George Washington University,
Washington, D.C. 20052}

\date{\today}
 
\begin{abstract}
 
The Chew-Mandelstam parameterization, which has been used extensively
in the two-body hadronic sector, is generalized in this exploratory
study to the electromagnetic sector by simultaneous fits to the
$\pi$-- and $\eta$--photoproduction $S$--wave multipole amplitudes for
center-of-mass energies from the pion threshold through 1.61 GeV. We
review the Chew-Mandelstam parameterization in detail to clarify the
theoretical content of the \said\ hadronic amplitude analysis and to
place the proposed, generalized \said\ electromagnetic amplitudes in
the context of earlier employed parameterized forms. The
parameterization is unitary at the two-body level, employing four
hadronic channels and the $\gamma N$ electromagnetic channel. 
We compare the resulting fit to the \maid\ parameterization and find
qualitative agreement though, numerically, the solution is somewhat
different.  Applications of the extended parameterization to global
fits of the photoproduction data and to global fits of the combined
hadronic and photoproduction data are discussed.
\end{abstract}


\maketitle

\section{Introduction}
\label{sec:intro}
Most of our knowledge of the excited baryons has come from fits to
hadronic scattering data, in particular pion-nucleon scattering, for
which there exists an accurate and nearly complete database extending
through and above the resonance region. Sufficient 
polarization observables exist to constitute complete measurements
over a significant kinematic interval.  The range of
hadroproduction data, including $\pn\to\pn$, $\pn\to\en$, $\pn\to\on$,
and other inelastic processes including, for example, strangeness 
production, have also
been used to constrain theoretical models and phenomenological 
parameterizations of the scattering and reaction amplitudes.

Currently, however, a renaissance is underway in meson production
and resonance physics with reaction data issuing from a number of
precision electromagnetic facilities. Collaborative theoretical 
and phenomenological efforts have started
to analyze these data in ways consistent with some subset
of constraints imposed by quantum field theory upon the reaction
amplitudes. The
quality and quantity of data in electromagnetic induced reactions
is becoming sufficient to rival and possibly surpass the hadroproduction
data. Since the electromagnetic reactions proceed mainly through 
the hadronic channels, the new data offers the possibility of
``back-constraining'' the hadronic amplitudes, conventionally determined
only in fits to the hadroproduction data.

It is in this context that we have completed an exploratory study of
the $S$--wave $\pi$-- and $\eta$--photoproduction multipoles in the
``Chew-Mandelstam'' approach, related to the $N/D$ representation, to
the electromagnetic reaction amplitude. The novel concept, which
provokes and permits this exploratory study, is the generalization of
the Chew-Mandelstam approach to the electromagnetic sector. We have
developed a new form for the amplitude that incorporates multichannel
hadronic rescattering effects in a complete manner consistent
with unitarity.
The near-term objective is to develop a framework in which to 
analyze the hadro- and electroproduction reactions simultaneously 
in a global framework.

Recent experimental observations of the photoproduction of the $\eta$
meson from the proton have yielded measurements of the unpolarized
differential cross section \cite{Crede:2003ax,Nakabayashi:2006ut,
Williams:2009yj}
and photon beam asymmetry\cite{Ajaka:1998zi,Elsner:2007hm} of high 
precision. Forthcoming measurements from the CLAS Collaboration at
Jefferson Lab\cite{Dugger:2010etaprop} and Mainz\cite{McGeorge:2007tg}
will rival, if not surpass, the precision of the existing measurements.

Several interesting features of $\eta$ meson physics motivate these measurements
and their theoretical interpretation in various
fields of nuclear physics, astrophysics, and particle physics. The possibility
that the $\eta$-nucleon interaction may be 
attractive\cite{Cheng:1987yw,Haider:2009yf}
suggests the existence of bound states of the $\eta$ meson with nuclei. Certain
resonances, the $S_{11}(1535)$ $N^*$ resonance in particular, 
are significantly coupled to the $\eta N$ channel, and the photoproduction
of this final state provides an independent method to probe the 
isospin $T=\shalf$ resonance spectrum and its 
couplings\cite{Prakhov:2005qb}.

The strong interactions of the $\pi$ and $\eta$ 
mesons require multichannel descriptions 
which respect unitarity in the relevant channel space in order to obtain a
realistic description of the data. The Chew-Mandelstam $K$-matrix 
approach \cite{Basdevant:1978tx,Edwards:1980sa,Arndt:1985vj} is an effective
parameterization of the observed reaction data since the elements of the 
Chew-Mandelstam $K$ matrix may be assumed to be real if the couplings 
to neglected, open channels are small.

Several relatively recent $K$ matrix analyses of the coupled 
$\pn$, $\en$, and $\gn$
channels\cite{Green:1997yia,Arndt:1998nm,Green:1999iq,Crede:2009zz} 
have been successful in obtaining reasonable parameterizations of the 
two-body partial wave amplitudes\cite{Green:note}. The
purpose of the present work is to investigate the extent to which a
description of the $\pi$ photoproduction $E^{1/2}_{0+}(S_{11})$ amplitude 
and the modulus of the $\eta$ photoproduction amplitude yields an 
$\eta$ photoproduction multipole with a resonant phase. Various 
calculations\cite{Kaiser:1996js,Tiator:1998qp,Green:1999iq,Aznauryan:2003zg},
indicate that the modulus of the $\eta$ photoproduction amplitude 
near threshold is fairly model-independent, being reproduced in a range 
of calculational models or schemes. In the present work, we take as input
hadronic $T$ matrix elements, determined in realistic ($\chi$-squared 
per datum $\sim 1$) fits to data \cite{Arndt:2006bf}, as discussed in
the following sections.

In Sec.\ \ref{sec:formal}, we review in some detail the
Chew-Mandelstam form\cite{Arndt:1985vj} of the parameterization.
The purpose of this review is to establish the theoretical 
considerations that motivate the amplitude parameterizations 
used in the \said\ program, to place these amplitudes in the 
context of other hadronic amplitude parameterization schemes and to 
lay the groundwork for future improvements.
Section \ref{sec:results} gives the results for the fits to the
isospin $T=\shalf$ $\pi$--photoproduction amplitude, $\Ezp$, 
and the modulus of the $\eta$ photoproduction
amplitude, $\Eze$. The conclusions are given in Sec.\ \ref{sec:conclusion}.
We find, in this exploratory study, an $\eta$ photoproduction
multipole having a resonant shape, qualitatively similar to a
Breit-Wigner form, and similar to other 
calculations\cite{Kaiser:1996js,Chiang:2001as,Aznauryan:2003zg}.
There is, however, significant deviation from the simple 
Breit-Wigner form.

\section{Chew-Mandelstam parameterization}
\label{sec:formal}
Previous work in the determination of the $\eta$ photoproduction
amplitudes\cite{Green:1997yia,Arndt:1998nm,Green:1999iq} has shown 
that an approach
which includes the coupling of the electromagnetic channel to the
$\pn$ and $\en$ channels in the region of energies near the center-of-mass
energy, $W=1535$ MeV gives a reasonably good description of the data and
a plausible form for the amplitudes. However, as our ultimate
objective is the simultaneous parameterization of hadro- and 
photoproduction scattering and reaction observables, we will go beyond
the two-channel treatment for this study of the $\Eze$ multipole amplitude.

\subsection{Unitarity constraint}
\label{subsec:uc}
The form of the Chew-Mandelstam parameterization, which we employ
in this study follows as a consequence of the analytic structure 
imposed by the 
unitarity\cite{Zimmerman:1961aa,Eden:1952aa,Polkinghorne:1962aa,
Boyling:1964aa,Bjorken:1960zz}
of the $S$ matrix in the physical region, $W>m_i+m_t$, 
where $W$ is the center-of-mass energy and $m_i$ and $m_t$ are
the masses of the incident and target particles. 
Confining our attention to two-particle initial and final states,
the $S$ matrix is defined as
\begin{align}
S_{\alpha\beta}(E) &= 
\bra{\kvec_\alpha \alpha}S\ket{\kvec_\beta \beta} \\
\label{eqn:Sdef}
&= \dth{\kvec_\alpha-\kvec_\beta}\delta_{\alpha\beta} \nonumber \\
&+ 2i \pi\delta(E_\alpha-E_\beta)
\bra{\kvec_\alpha \alpha}T\ket{\kvec_\beta \beta}
\end{align}
where $\kvec_{\alpha,\beta}$ are the final and initial relative
momenta, respectively, $E=E_\alpha=E_\beta=W$ is the 
center-of-mass energy, and
the labels $\alpha$ and $\beta$ denote the particle species, spins,
and internal quantum numbers, such as isospin.
The initial and final energies, $E_\beta$ and $E_\alpha$, respectively
are related 
to the on-shell relative momenta for channel $\alpha$, $\kbar_\alpha$ as
\begin{align}
W &= E_{\alpha,1} + E_{\alpha,2} \\
E_{\alpha,i} &= \sqrt{\bar k_\alpha^2 + m_{\alpha,i}^2}.
\end{align}
The on-shell relative momentum may be expressed in terms of
the center-of-mass energy, $W$, as
\begin{align}
\label{eqn:on-shellk}
\bar k_\alpha &= 
\frac{1}{2W} \sqrt{W-m_{\alpha +}} \sqrt{W-m_{\alpha -}} \nonumber \\
&\times      \sqrt{W+m_{\alpha +}} \sqrt{W+m_{\alpha -}},
\end{align}
with $m_{\alpha\pm} = m_{\alpha,1}\pm m_{\alpha,2}$.

The scattering operator, $S$, is unitary
\begin{align}
S^\dag S &= S S^\dag = 1
\end{align}
and if we restrict our analysis to energies where just
two-particle channels contribute, we obtain
\begin{align}
&\sum_\sigma \intks
\bra{\kvec_\alpha \alpha}S^\dag\ket{\kvec_\sigma \sigma}
\bra{\kvec_\sigma \sigma}S\ket{\kvec_\beta \beta} \nonumber \\
&= \dth{\kvec_\alpha-\kvec_\beta}\delta_{\alpha\beta}.
\end{align}
Substitution of Eq.\eqref{eqn:Sdef} into the relation above yields the
unitarity constraint on $T$
\begin{align}
\label{eqn:UCT1}
T_{\alpha\beta} - T^\dag_{\alpha\beta}
&= 2\pi i \sum_\sigma \intks T^\dag_{\alpha\sigma}
\delta(E_\alpha-E_\sigma)
T_{\sigma\beta}.
\end{align}
Effecting the integration on $k_\sigma\equiv|\kvec_\sigma|$ gives
\begin{align}
\label{eqn:uct-thr}
 T_{\alpha\beta} - T^\dag_{\alpha\beta}
&= 2i \sum_\sigma 
\int\!\! d\Omega_\sigma 
T^\dag_{\alpha\sigma} \theta(W-m_{\sigma+})\rho_{\sigma} T_{\sigma\beta}
\end{align}
where
\begin{align}
\rho_\sigma(\bar k_\sigma) &=
\frac{\pi \bar k_\sigma E_{\sigma 1}E_{\sigma 2}}{W}.
\end{align}
The presence of the Heaviside step function, $\theta(W-m_{\sigma+})$
is a consequence of the fact that, over the range of integration
$k_\sigma>0$, the argument of the $\delta$ function has a solution, 
$E_\alpha-E_\sigma=0$ only when $W>m_{\sigma+}$.
Equation \eqref{eqn:uct-thr} implies discontinuities
in the derivative of the imaginary part at each channel threshold
$W=m_{\sigma +}$:
\begin{align}
\frac{1}{2i}[T_{\alpha\beta} - T^*_{\alpha\beta}]
&= \mbox{Im } T_{\alpha\beta} \\
&= \sum_\sigma 
\int\!\! d\Omega_\sigma 
T^*_{\alpha\sigma} \theta(W-m_{\sigma+})\rho_{\sigma} T_{\sigma\beta}
\end{align}
where we have assumed that, due to the time-reversal invariance of the
strong interaction, $T_{\alpha\beta}=T_{\beta\alpha}$.
The violation of the Cauchy-Riemann equations at threshold indicates
the presence of a branch point. We distinguish between the dynamical
singularities at each threshold opening $m_{\sigma+}$ and kinematical
singularities, due to the presence of kinematical factors such as
$\kbar_\sigma$. The kinematical singularities are removed from the
unitarity constraint by considering $T'_{\alpha\beta} = \sqrt\rho_\alpha
T_{\alpha\beta} \sqrt\rho_\beta$.

We may transform to the partial
wave representation 
and write
\begin{align}
T'_{\alpha\beta} - T'^*_{\alpha\beta} 
&= 2i \sum_\sigma T'^*_{\alpha\sigma}
\theta(W-m_{\sigma+}) T'_{\sigma\beta}
\end{align}
where the $T'_{\alpha\beta}$ now represent the partial wave amplitudes.
Casting this relation as a matrix equation
\begin{align}
\frac{1}{2i} [T'-T'^*] &= T'^*\theta(W-M_+) T',
\end{align}
where $M_{+,\alpha\sigma} = m_{\sigma+}\delta_{\alpha\sigma}$,
and multiplying from the left by $[T'^*]^{-1}$ and from the right by
$T'^{-1}$ gives
\begin{align}
\label{eqn:UCTinv}
\mbox{Im } T'^{-1} &= -\theta(W-M_+),
\end{align}
a diagonal matrix. Since this equation isolates the imaginary part
of the inverse-$T$ matrix, we write
\begin{align}
T'^{-1} &= \mbox{Re } T'^{-1} + i\mbox{Im } T'^{-1}, \\
\label{eqn:Kinv}
       &= K'^{-1} - i\theta(W-M_+),
\end{align} 
where we've defined $\mbox{Re } T'^{-1} = K'^{-1}$ and 
$K_{\alpha\beta}'=\sqrt{\rho_\alpha} K_{\alpha\beta} \sqrt{\rho_\beta}$. 
Multiplying from one
side by $T'$ and the other by $K'$ gives the Heitler integral
equation\cite{Heitler:1941a,Goldberger:1964aa}
\begin{align}
\label{eqn:TpK}
T' &= K' + K' i\theta(W-M_+) T'.
\end{align}
This is the starting point for the Chew-Mandelstam parameterization of
the reaction amplitude.

We emphasize that, in the physical region, the unitarity relation is
satisfied by the imaginary part of $T'^{-1}$. Therefore the Heitler
$K$ matrix is analytic, except for possible isolated 
poles\cite{Workman:2008iv}, throughout the physical 
region\cite{Zimmerman:1961aa,Eden:1966sm}.
This is apparent if we consider a dynamical equation of, for example,
the Lippmann-Schwinger form:
\begin{align}
\label{eqn:TLS}
T &= V + V G_0 T, \\
\label{eqn:GLS}
G_0 &= \mathcal{P}\frac{1}{E-H_0} - i\pi\delta(E-H_0).
\end{align}
Here $V$ is the interaction part of the full Hamiltonian,
$E=W$, $H_0$ is the free-particle Hamiltonian, and $\mathcal{P}$
denotes the Cauchy principal value prescription. Substitution of
Eq.\eqref{eqn:GLS} into Eq.\eqref{eqn:TLS} gives $T=K+iK\delta(E-H_0)T$
where
\begin{align}
\label{eqn:LSK}
K &= V + V \mathcal{P}\frac{1}{E-H_0} K.
\end{align}
The Cauchy principal value prescription in this equation yields a
kernel which is completely continuous in the physical region. The
spectrum of the kernel therefore possesses no eigenvalues in the
continuum and $K$ is analytic (other than possible poles)
there\cite{Weinberg:1964zz}. The $K$ matrix may possess singularities
in other regions of the complex energy plane.  In fact, the
interaction $V$ possesses singularities in regions outside the
physical
region\cite{Mandelstam:1958xc,Frautschi:1960aa,Frazer:1960zz}. In
particular, there is a branch point at some value $W<0$. We intend to
neglect singularities in the region $\mbox{Re } W<0$ for the purposes
of the present study and avoid a detailed discussion of them here. A
description is available in the
literature\cite{Mandelstam:1958xc,Frautschi:1960aa,Frazer:1960zz,Chew:1960iv}.
Inclusion of singularities in the region $\mbox{Re } W<0$ will be
explored in subsequent investigations.

The partial wave amplitude is therefore 
known to have the following singularities.
There are branch points in the physical region at the channel-opening 
thresholds as in Eq.\eqref{eqn:uct-thr}, branch points in the region $W<0$,
and possible poles consistent with causality\cite{Eden:1966sm,Queen:1974dr}.
An efficient parameterization following Ref.\cite{Chew:1960iv}, which 
encodes these singularities, involves the factorization of the partial 
wave amplitude. This is referred to as the ``$N/D$'' approach. We will 
use the $N/D$ language to clarify the nature of the singularities of the 
$T$ matrix which are included and those neglected in our Chew-Mandelstam 
approach.

\subsection{Relation to $N/D$ approach}
The $N/D$ approach has been used to analyze a variety of 
reactions\cite{Chew:1960iv,Barut:1967ao,Arndt:1968sm}. 
As our long term objective
is the generalization of the existing method used to parameterize the
hadronic and electromagnetic amplitudes, we collect here some of the
relevant equations of the $N/D$ approach. The $T$ matrix is written in 
the factorized form
\begin{align}
T(W) &=  D^{-1}(W) N(W)
\end{align}
where $N$ and $D$ are $\nc\times\nc$ arrays\cite{Bjorken:1960zz},
where $\nc$ is the number of included two-body channels. 
This relation has been
shown to be consistent with the requirement of time-reversal invariance
in Ref.\cite{Bjorken:1961nd}. The relations
\begin{align}
\label{eqn:Dcut}
\im D(W) &= N(W) \im T^{-1}(W) & W>m_i+m_t \\
\im N(W) &= 0 & W>m_i+m_t \\
\label{eqn:Ncut}
\im N(W) &= D(W) \im T(W) & W<0 \\
\im D(W) &= 0 & W<0
\end{align}
give the essential content of the $N/D$ approach. They state that the
function $D$ has branch points only in the physical, $W>m_i+m_t$ region and 
that $N$ has only unphysical, $W<0$ branch points. These relations determine 
the following dispersion relation (or Hilbert transform) representation
for $D$
\begin{align}
D(W) &= \sum_{i=1}^{n_p}D(W;W_i) - \frac{1}{\pi}\prod_{i=1}^{n_p}(W-W_i)
\nonumber \\
&\times\int_{W_t}^\infty dW'\frac{N(W')\rho(W')}
{(W'-W)\prod_j(W'-W_j)},
\end{align}
with $n_p$ subtractions. Here, $D(W;W_i)$ is a polynomial of order $n_p$,
$W\in\C$, and
$W_t$ is the lowest production threshold. Here, we show the polynomial 
ambiguity of the Hilbert transform explicitly to allow for the possibility
that the parameterization includes several subtraction points.

Using the relation $T = ND^{-1}$ in the physical region,
the numerator factor, $N$ can be shown to satisfy the integral equation
\begin{align}
N(W) &= K\left\{\sum_i D(W;W_i) 
-\frac{1}{\pi}\prod_{i=1}^{n_p}(W-W_i)\right. \nonumber \\
&\left.\times
\dashint_{W_t}^\infty dW' \frac{N(W')\rho(W')}{(W'-W)\prod_j(W'-W_j)}\right\}
\end{align}
where $\dashint$ denotes the Cauchy principal value integral, and
the Heitler $K$ matrix, $K$ is defined by Eq.\eqref{eqn:TpK}, with 
$K=\rho^{-\half}K'\rho^{-\half}$.

\subsection{Chew-Mandelstam parameterization}
\label{subsec:CM}
The preceding discussion of unitarity and the $N/D$ approach provides
the context for our present parameterization.
The Chew-Mandelstam parameterization developed here is similar to 
those of Refs.\cite{Basdevant:1978tx,Edwards:1980sa} and \cite{Arndt:1985vj}. 
We consider Eq.\eqref{eqn:Kinv} and rewrite it, confining our attention
to the $S$ wave multipole as
\begin{align}
\label{eqn:TinvK}
T^{-1} &= K^{-1} - i \tilde\rho \\
&= (K^{-1} + \mbox{Re}\,C) - (\mbox{Re}\,C + i\tilde\rho) \nonumber \\
\label{eqn:TinvKbar}
 &= \Kbar^{-1} - C,
\end{align}
where $\tilde\rho = \rho\theta(W-M_+)$ and $\mbox{Im}\,C = \tilde\rho
= \theta(W-M_+)\rho$. The transition matrix is given in terms of the
``Chew-Mandelstam'' (CM) $K$ matrix, \Kbar\ by
\begin{align}
\label{eqn:TKbar}
T &= \Kbar + \Kbar C T.
\end{align}

Equation \eqref{eqn:TKbar} fixes our Chew-Mandelstam parameterization. In
the language of the preceding section, we have neglected the $W<0$ branch
points of $N$ and made the approximation $N(W)=\Kbar(W)$, an entire function. 
The ``Chew-Mandelstam'' function, $C_\alpha$ is determined solely by the 
unitarity constraint, Eq.\eqref{eqn:UCTinv} since Eq.\eqref{eqn:TinvKbar}
is equivalent to taking $D=1-\Kbar C$. Then the Chew-Mandelstam function
 is given by a Cauchy integral over the discontinuity of $C_\alpha$ in
the physical region
\begin{align}
\label{eqn:CMfunc}
C_\alpha(W) &= \int_{W_t}^\infty \frac{dW'}{\pi} \frac{\rho_\alpha(W')}{W'-W}
\nonumber \\
            &- \int_{W_t}^\infty \frac{dW'}{\pi} \frac{\rho_\alpha(W')}{W'-W_s},
\end{align}
where we have made one subtraction, $0 \le W_s < W_t$. 
Defining $\zbar_\alpha=\frac{W-W_{t,\alpha}}{W-W_{s,\alpha}}$ we can 
rewrite Eq.\eqref{eqn:CMfunc} as
\begin{align}
C_\alpha(W) &= \int_0^1 \frac{dx}{\pi} \frac{\rho(x)}{x-\zbar_\alpha(W)}.
\end{align}

The relationship between the Heitler
$K$ matrix and the CM $K$ matrix, $\Kbar$ is given by
\begin{align}
\label{eqn:KKbar}
K &= \Kbar + \Kbar [\mbox{Re}\, C] K.
\end{align}
This demonstrates a possible advantage of using the CM $K$ matrix. If
we consider a polynomial parameterization of a given CM $K$ matrix element
\begin{align}
\label{eqn:Kbarpar}
\Kbar_{\alpha\beta} &= \sum_{n=0}^{n_{\alpha\beta}}
c_{\alpha\beta,n} \zbar^n_{\alpha\beta}
\end{align}
where $n_{\alpha\beta}$ are channel dependent integers controlling the order
of the polynomial (polynomials typically less than fifth order are used) and
$\zbar_{\alpha\beta}$ is a possibly channel dependent linear function of the
center-of-mass energy, $W$ then we see, by solving Eq.\eqref{eqn:KKbar} for
$K$
\begin{align}
K &= \frac{1}{1-\Kbar [\mbox{Re}\,C]} \Kbar
\end{align}
that poles may appear in the $K$
matrix. Attempts to relate the $K$ matrix poles to resonances have been
made\cite{Martin:1970bb,Davidson:1990yk,Ceci:2006jj}. Here, we simply 
point out that, though $K$ matrix poles are not simply related to
$T$ matrix poles\cite{Workman:2008iv}, Eq.\eqref{eqn:KKbar} shows that
one need not explicitly include pole terms in $\Kbar$ in order
to have poles in $K$.
Parameterizing $\Kbar(W)=N(W)$ as a polynomial, as noted, neglects
singularities in the unphysical region, $W<0$\cite{Adler:1964zero}. 
The branch points there
and discontinuities across their associated branch cuts are determined
by the production mechanisms\cite{Babelon:1976kv} 
relevant for the reaction considered. 

There are at least two reasons why polynomials may provide a reasonable 
starting point for a realistic parameterization of multichannel scattering 
and reaction amplitudes. The unitarity branch points, given their
physical nature, largely determine the gross structure of the
amplitudes in the physical region. This leads in an obvious way 
to the supposition that
more distant singularities in the complex $W$ plane associated, in 
particular, with the branch points in the unphysical region may be
less important. Experience has also confirmed this to be true.
The existing \said\ parameterizations of $\pn$ elastic 
scattering\cite{Arndt:2006bf}, 
the $\pn\to\en$\cite{Prakhov:2005qb} reaction, 
$\pi$--photoproduction\cite{Arndt:1989ww,Arndt:1990ej}
and electroproduction\cite{Arndt:2001si} and other reactions
all reveal that a realistic description with
$\chi^2$ per datum in the range of $1-3$ is possible with the polynomial
approximation for $\Kbar$.

\begin{figure}[b]
\includegraphics[width=250pt,keepaspectratio,clip]{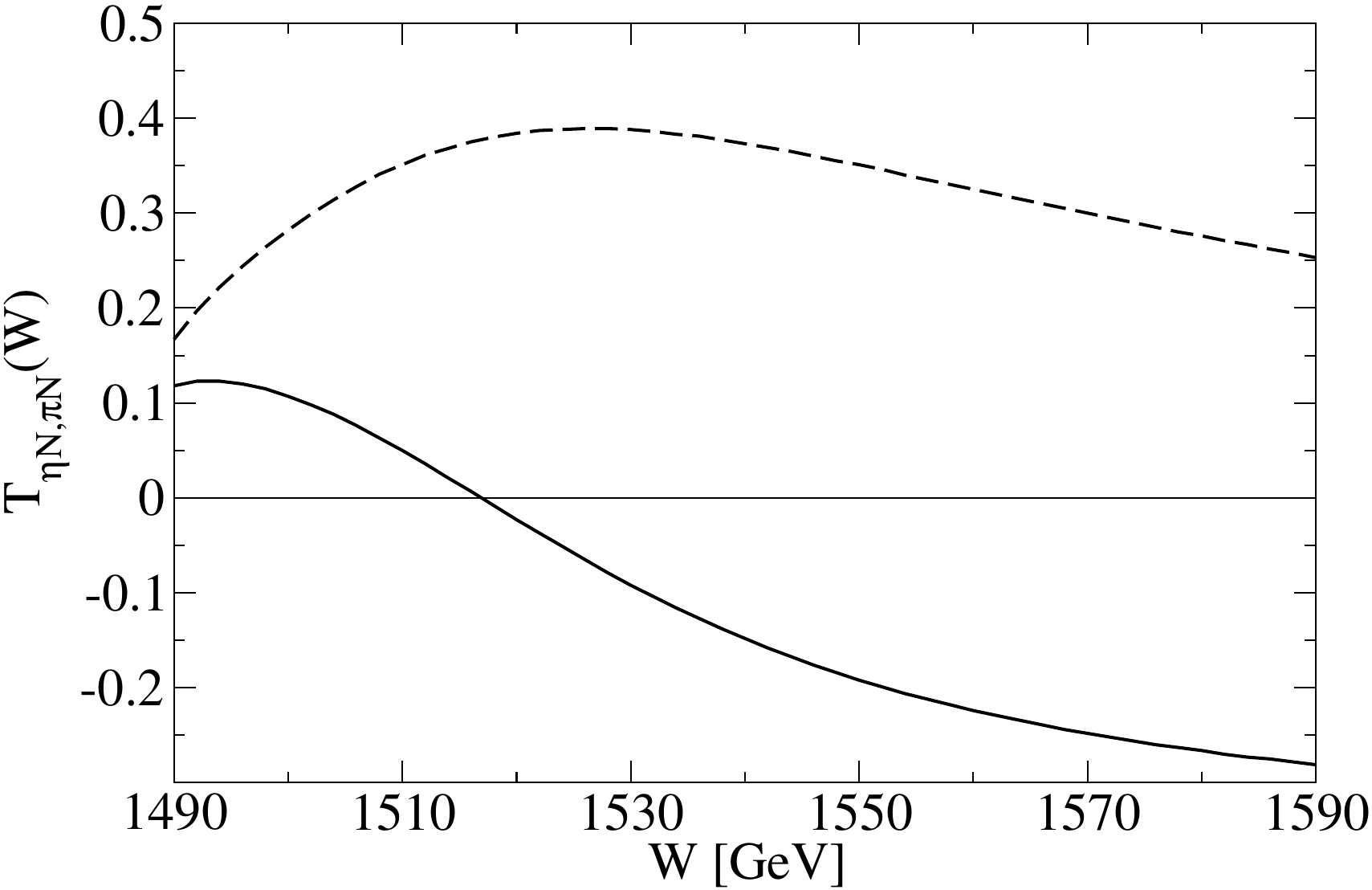}
\caption{\label{fig:pnenS11}The \said\ $S_{11}$ multipole for the $\pn\to\en$ 
reaction as a function of energy, $W$\cite{Prakhov:2005qb}. The solid
(dashed) line is the real (imaginary) part of the amplitude.}
\end{figure}

\begin{table}
\begin{center}
\newcolumntype{d}[1]{D{.}{.}{#1}}
\begin{tabular*}{0.75\textwidth}{@{\extracolsep{\fill}}l|d{1}d{1}|d{1}d{1}|d{1}d{1}|d{1}d{1}|d{1}d{1}}
 &\multicolumn{2}{c}{{\sc SP06}} &\multicolumn{2}{c}{{\sc FA02}} &\multicolumn{2}{c}{{\sc KA84}}&\multicolumn{2}{c}{{\sc EBAC}}&\multicolumn{2}{c}{{\sc Giessen}} \\
 \hline
$\pi^+p\to\pi^+p$ &
2.0 & 6.1 & 2.1 & 8.8 & 5.0 & 24.9 & 13.1 & 23.7 & 10.5 & 17.7 \\ 
$\pi^-p\to\pi^-p$ & 
1.9 & 6.2 & 2.0 & 6.6 & 9.1 & 51.9 &  4.9 & 16.0 & 12.1 & 34.1 \\
$\pi^-p\to\pi^0n$ & 
2.0 & 4.0 & 1.9 & 5.9 & 4.4 &  8.8 &  3.5 &  6.3 &  6.3 & 15.2 \\
$\pi^-p\to\eta n$ &
2.5 & 9.6 & 2.5 & 10.5 &-&-&-&-&-&-
\end{tabular*}
\end{center}
\caption{\label{tab:chi2}Normalized (left of each column pair) and 
unnormalized (right of each column pair) $\chi^2$-per-datum for
the {\sc SP06}\cite{Arndt:2006bf} and {\sc FA02}\cite{Arndt:2003if} solutions
of \said, {\sc KA84}\cite{Koch:1985bp}, EBAC\cite{JuliaDiaz:2007kz}, and
Giessen\cite{Shklyar:2004ba}. The energy ranges of the four groups are
from threshold to 2.5, 2.9, 1.91, and 2.0 GeV,
respectively.\cite{Tablenote}}
\end{table}

\begin{figure}
\includegraphics[width=250pt,keepaspectratio,clip]{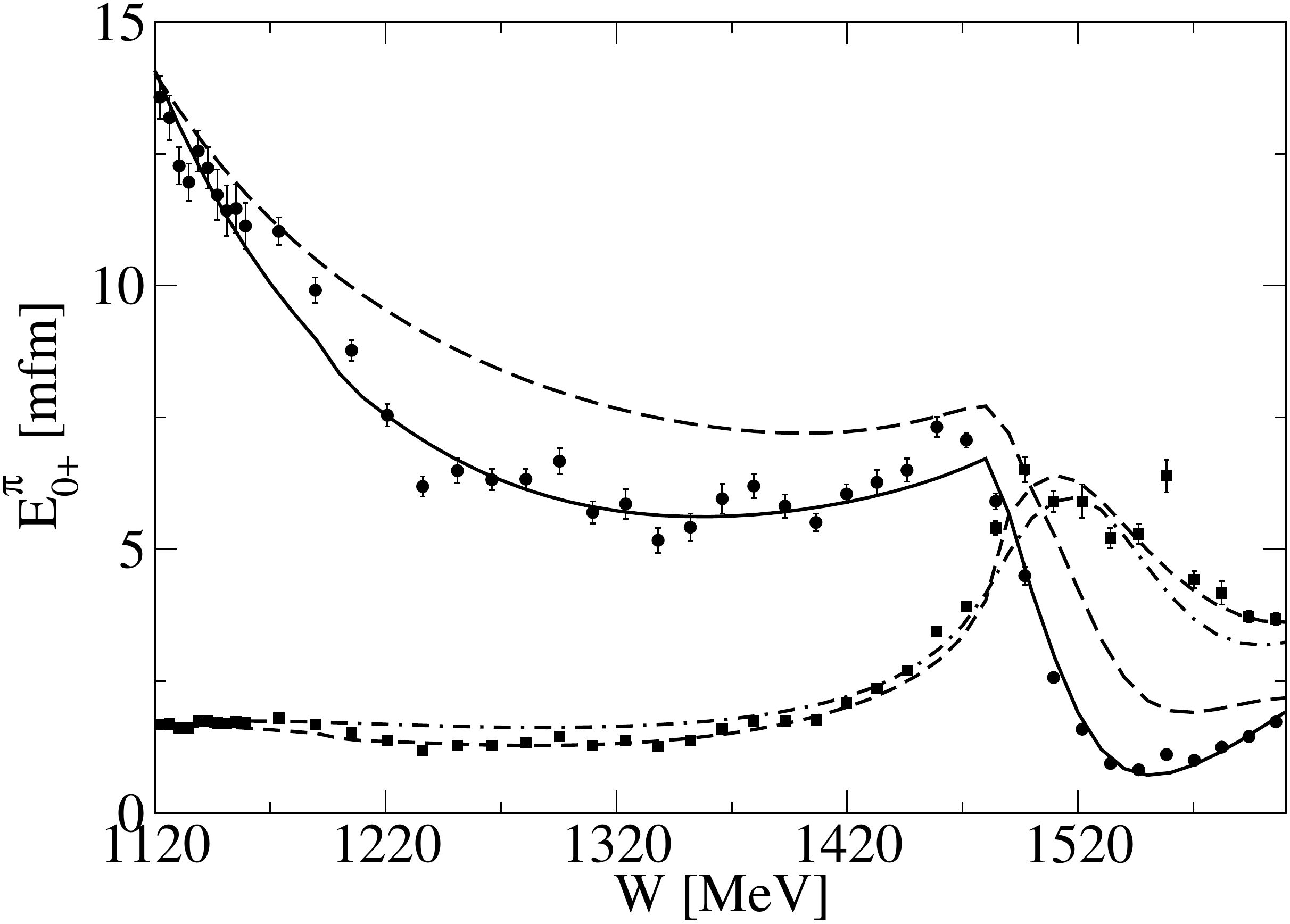}
\caption{\label{fig:picf}Comparison of the \said\ real (solid curve)
and imaginary (short-dashed curve) parts of the \EpiS\ multipole
amplitude with that of the \maid\cite{Drechsel:2007if} real (long-dashed
curve) and imaginary (dot-dashed curve) parts. The amplitudes are plotted
along with the real (circles) and imaginary (squares) of the 
\said\ single-energy solutions\cite{Arndt:2006bf}.}
\end{figure}

\section{Results}
\label{sec:results}
The Chew-Mandelstam parameterization for the $T$ matrix,
described in the preceding section, has been applied
recently\cite{Arndt:2006bf} to a coupled-channel 
fit for the $\pn$ elastic scattering and $\pn\to\en$ reaction.
It gives a realistic description of the data with $\chi^2$ per
datum better than any other parameterization or model, to our
knowledge. The $\chi^2$ per datum is shown in Table \ref{tab:chi2} 
against other parameterizations and model calculations for which
we possess sufficient amplitude information to perform such an 
analysis\cite{Tablenote}. 
The current \said\ parameterization 
used in this fit is given as
\begin{align}
\label{eqn:Tab}
T_{\alpha\beta} &= \sum_\sigma [1-\Kbar C]^{-1}_{\alpha\sigma}
\Kbar_{\sigma\beta}
\end{align}
where $\alpha,\beta$ and $\sigma$ are channel indices for the 
considered channels,
$\pn,\pD,\rn$ and $\en$. This parameterization has been discussed
in Refs.\cite{Arndt:1985vj,Arndt:1995bj,Arndt:2006bf}.
Given the success of this approach
in the hadronic two-body sector, the application to the study
of meson photoproduction is warranted.

The central result of the current exploratory study is to show that this
form can be extended to include the electromagnetic channel, 
\begin{align}
\label{eqn:Tag}
T_{\alpha\gamma} &= \sum_\sigma [1-\Kbar C]^{-1}_{\alpha\sigma}
\Kbar_{\sigma\gamma}
\end{align}
where $\gamma$ denotes the electromagnetic channel, $\gn$. Note that
Eqs.\eqref{eqn:Tab} and \eqref{eqn:Tag} share the common factor,
$[1-\Kbar C]_{\alpha\sigma}^{-1}$ which encodes, at least 
qualitatively speaking, the hadronic channel coupling 
(or rescattering) effects.

\begin{figure}[t]
\includegraphics[width=250pt,keepaspectratio,clip]{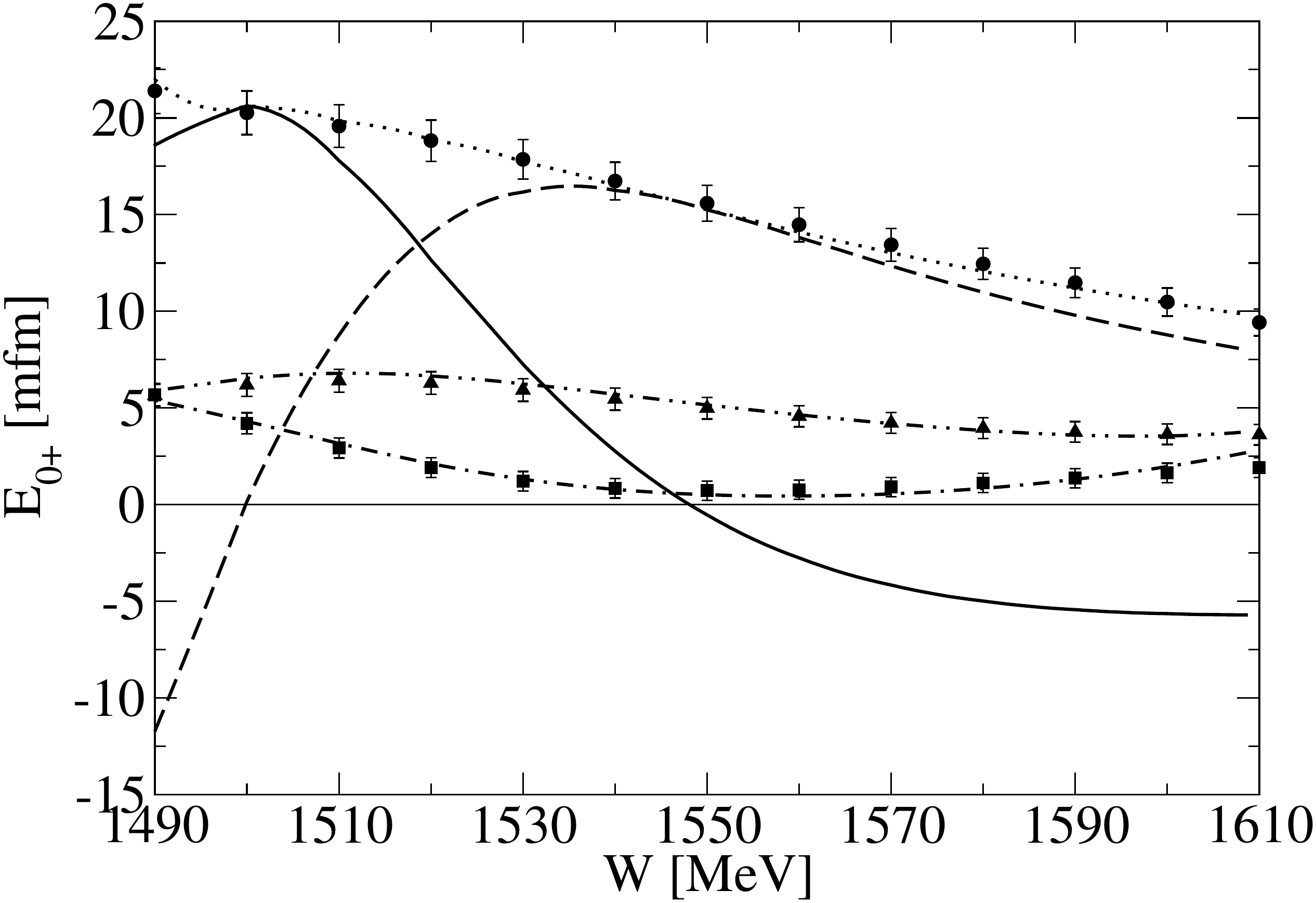}
\caption{\label{fig:Esaid}The predicted values for the real (solid curve) and
imaginary (dashed curve) for \EetaS versus the energy, $W$. 
The modulus $|\EetaS|$ (dotted curve),
the real (dot-dashed curve), and the imaginary (double
dot-dashed curve) parts of the $\pi$--photoproduction, $\EpiS$ were fit
to pseudodata generated from the \said\ solution\cite{Arndt:1989ww} with
the parameterized form Eq.\eqref{eqn:Tag} using 8 parameters (see text).}
\end{figure}

The form Eq.\eqref{eqn:Tag} for photoproduction should be contrasted 
with that currently employed in the $\pi$--photoproduction studies of 
Refs.\cite{Arndt:1989ww,Arndt:1990ej,Arndt:1998nm}
\begin{align}
\label{eqn:Tgamma-old}
T_{\pi\gamma} &= A(W) (1+iT_{\pi\pi}(W)) + iB(W)T_{\pi\pi}(W)
\end{align}
where the ``structure functions'' $A(W)$ and $B(W)$ are parameterized
as polynomials in the energy, $W$, $T_{\pi\gamma}
=T_{\pn,\gn}$ and $T_{\pi\pi}=T_{\pn,\pn}$, and the factor $A(W)$
contains a contribution from tree-level Born diagrams. This satisfies
Watson's theorem\cite{Watson:1952ji} (as does Eq.\eqref{eqn:Tag}), 
and is derived via the considerations
discussed in Ref.\cite{Workman:2005eu}.
While resulting in a realistic description of the data and being
comparable, at least qualitatively, with other parameterizations such
as \maid\cite{Chiang:2001as} for $\pi$--photoproduction, 
it does not satisfy 
the full multichannel unitarity constraint imposed by Eq.\eqref{eqn:uct-thr}.
This deficiency led us to consider the form in Eq.\eqref{eqn:Tag}, which 
manifestly satisfies the multichannel unitarity constraint, 
Eq.\eqref{eqn:uct-thr}.

\begin{figure}
\includegraphics[width=250pt,keepaspectratio,clip]{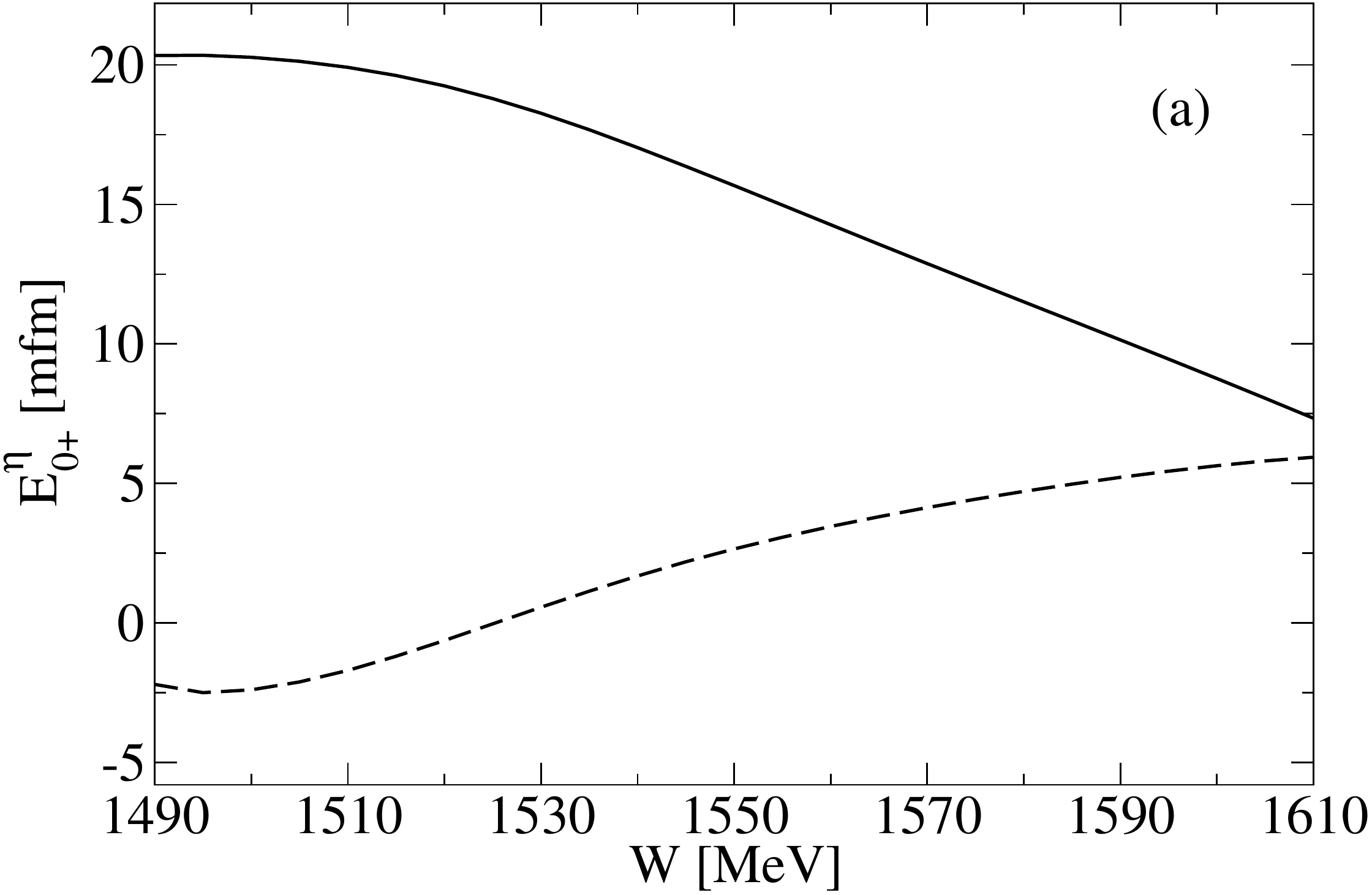}
\caption{\label{fig:E429}The $\eta-$photoproduction $S_{11}$ multipole
amplitude, \EetaS versus the energy, $W$ fit using the previously
employed, non-unitary form of Eq.\eqref{eqn:Tgamma-old}. The behavior
near $W\simeq 1535$ MeV is not resonant as can be clearly seen in
Fig.\eqref{fig:Argand-eta}.}
\end{figure}

\begin{figure}[b]
\includegraphics[width=250pt,keepaspectratio,clip]{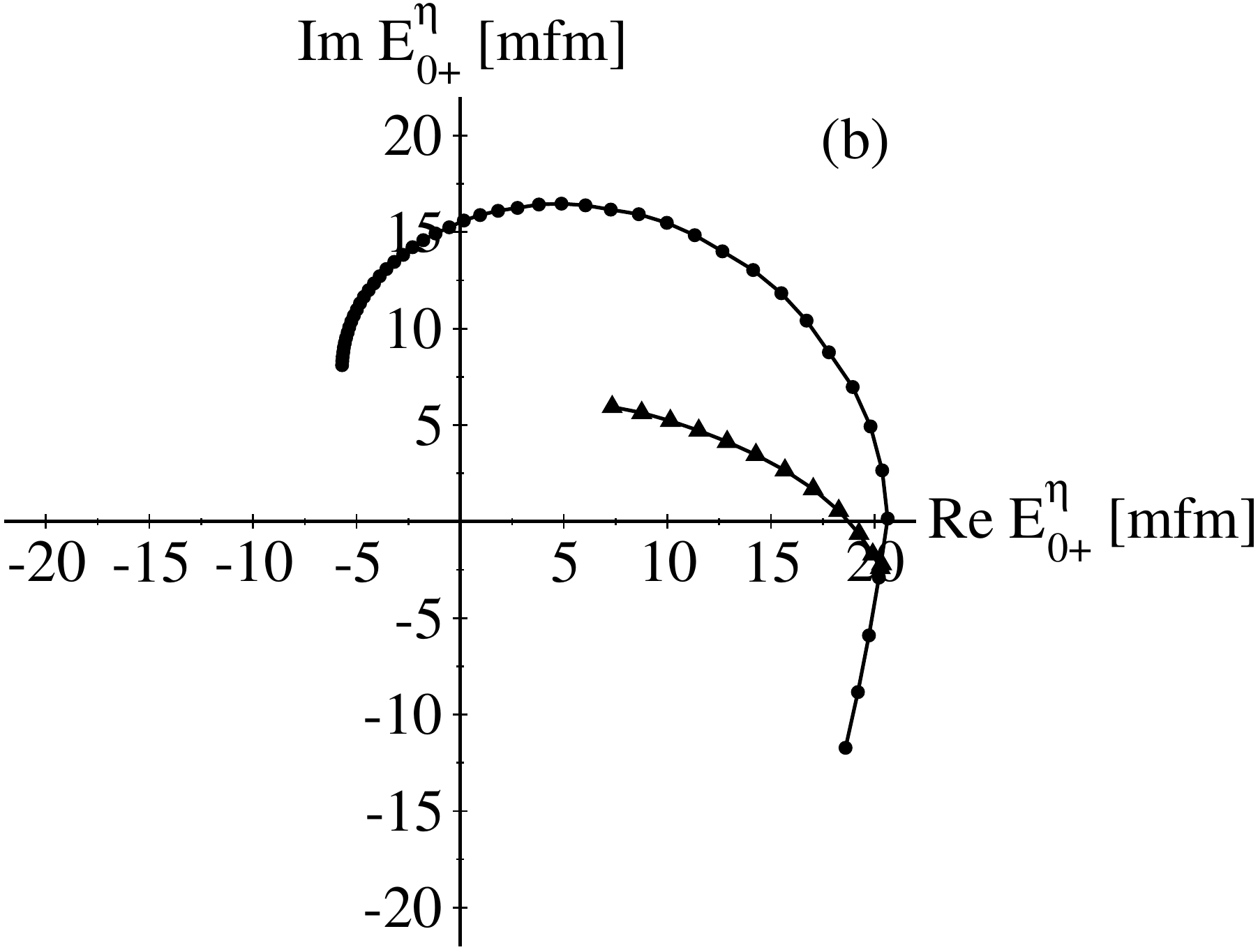}
\caption{\label{fig:Argand-eta}Argand-plot comparison of the
$\eta-$photoproduction $S_{11}$ multipole amplitudes, \re\EetaS\
versus \im\EetaS\ plotted in the range $1490$ MeV $\le W \le 1610$ MeV
of center-of-mass energy, $W$ with two fit forms. The curve with
energies marked by triangles is another representation of the result
for \EetaS shown in Fig.\eqref{fig:E429}, determined using the
parameterization of
Eq.\eqref{eqn:Tgamma-old}. The curve with energies marked by circles
is another representation of the result for \EetaS, shown in
Fig.\eqref{fig:Esaid}, determined using the parameterization
of Eq.\eqref{eqn:Tag}. The curves span the same interval in
energy but with different spacings. The first curve (triangles) is
clearly non-resonant in the region shown while the
second curve (cirlces) clearly shows resonant behavior; the apex on
the Argand diagram of the second curve occurs at precisely $W=1535$
MeV.}
\end{figure}

The need to include the multichannel unitarity effects of
Eq.\eqref{eqn:Tag} have also become apparent in difficulties faced in
attempts to parameterize the $\eta$--photoproduction reaction using
forms\cite{Arndt:2009pc} similar to Eq.\eqref{eqn:Tgamma-old}. Forms
of this type, used in fits to the $\eta$--photoproduction data alone,
yielded an $S$--wave multipole without a clearly resonant shape, even
while yielding fits to the observed data with realistic $\chi^2$ per
datum on the order of 2 to 4. An example of such a fit employing
Eq.\eqref{eqn:Tgamma-old} is shown in Fig.\eqref{fig:E429}.  Near
values of the center-of-mass energy $W\simeq 1535$ the amplitude in
Fig.\eqref{fig:E429} is decidedly not resonant. This is also clear in
the Argand plot of Fig.\eqref{fig:Argand-eta}. Here we have shown the
comparison of the fit forms used in Ref.\cite{Arndt:2009pc} (with
energies marked by triangles) This difficulty was an early motivation
for the present study.  Expectation of resonant behavior for
$\eta$--photoproduction, $\gn\to\en$ in the $S$--wave can be argued
straightforwardly. For example, since the electromagnetic coupling to
the $\pn$ channel is large, the $\gn\to\en$ reaction may proceed via
the $\pn\to\en$ amplitude of Fig.\eqref{fig:pnenS11}, or through
direct resonance production. We therefore anticipate the hadronic
subprocess will drive a significant resonant effect in the (isoscalar)
electromagnetic transition.

Several other studies have determined that a resonant structure near
$W \sim 1535$ MeV is consistent with the reaction data. These works
include those in
Refs.\cite{Kaiser:1996js,Green:1999iq,Chiang:2001as,Aznauryan:2003zg}.
We should note that all of these works have assumed the $S_{11}$ wave
to be resonant, usually by including a Breit-Wigner or similar term
explicitly into their formalism. We do not make this assumption in
using Eq.\eqref{eqn:Tag}. 

\begin{figure}[t]
\includegraphics[width=250pt,keepaspectratio,clip]{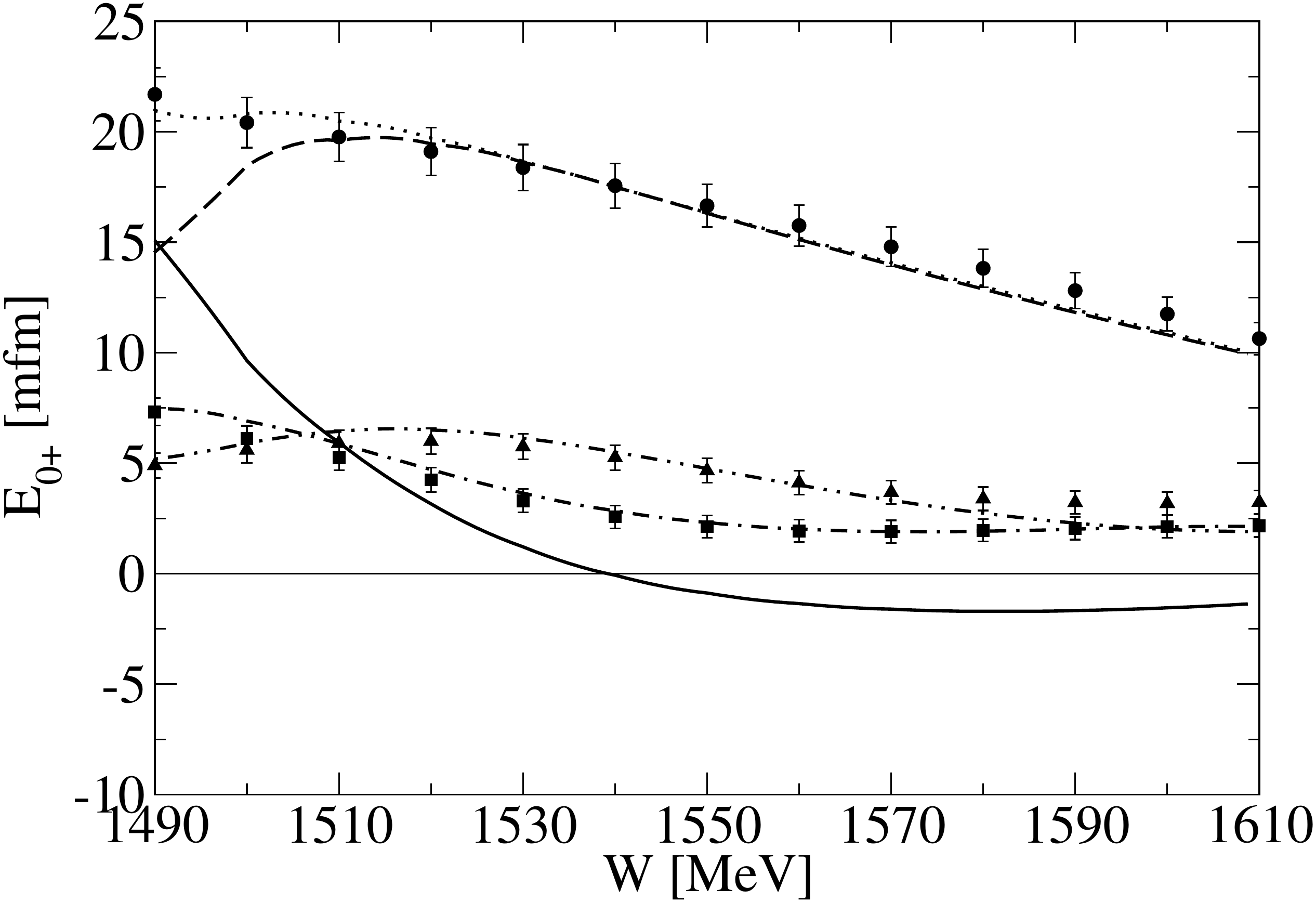}
\caption{\label{fig:E0+MAID-7}The predicted values for the real 
(solid curve) and imaginary (dashed curve) for \EetaS versus the
energy, $W$. The modulus $|\EetaS|$ (dotted curve),
the real (dot-dashed curve), and the imaginary (double
dot-dashed curve) parts of the $\pi$--photoproduction, $\EpiS$ were fit
to pseudodata generated from the \maid\ solution\cite{Chiang:2002vq} with 
the parameterized form Eq.\eqref{eqn:Tag} using 7 parameters (see text).}
\end{figure}

\begin{figure}[b]
\includegraphics[width=250pt,keepaspectratio,clip]{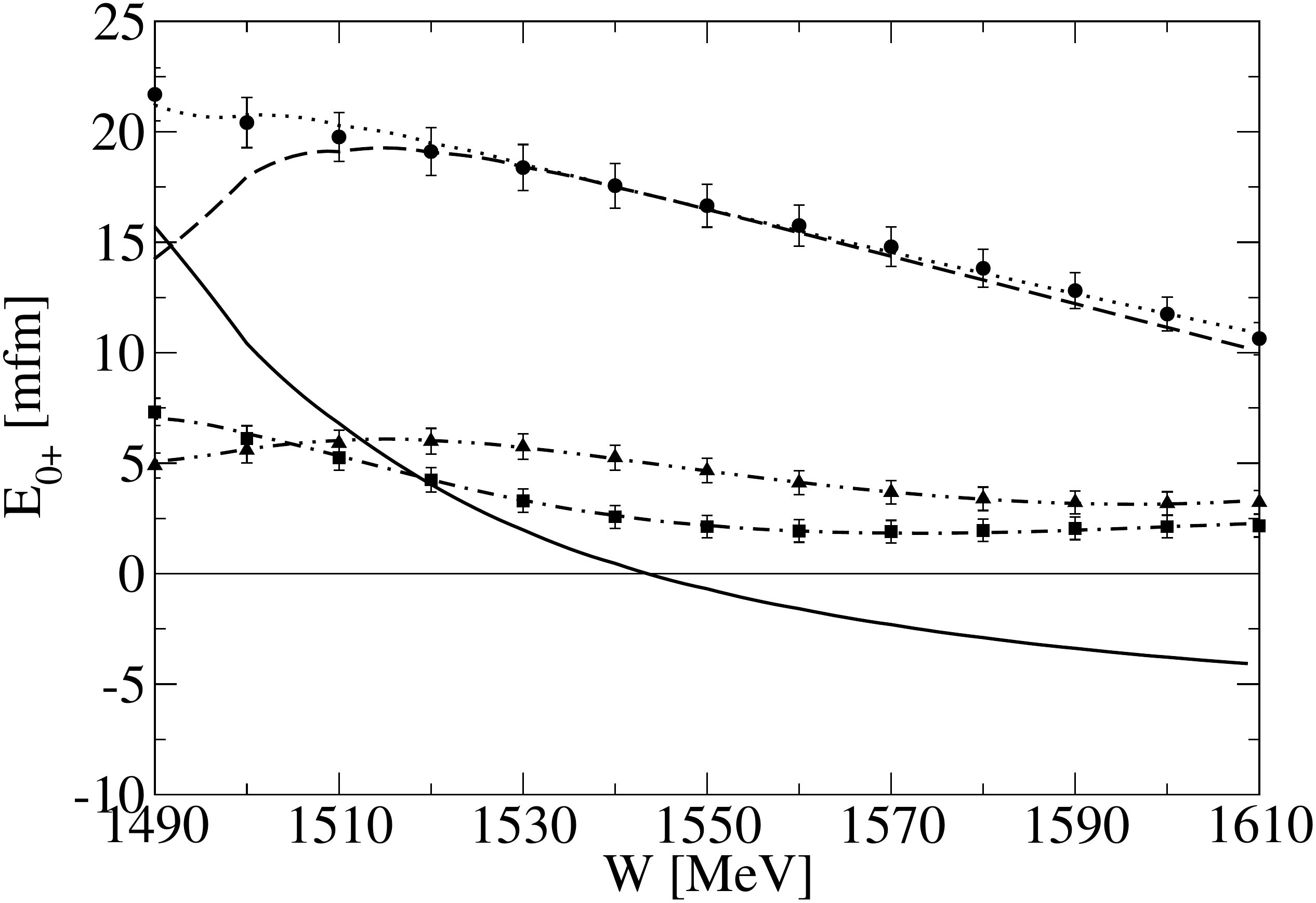}
\caption{\label{fig:E0+MAID-14}The predicted values for the real 
(solid curve) and imaginary (dashed curve) for \EetaS versus the
energy, $W$. The modulus $|\EetaS|$ (dotted curve),
the real (dot-dashed curve), and the imaginary (double
dot-dashed curve) parts of the $\pi$--photoproduction, $\EpiS$ were fit
to pseudodata generated from the \maid\ solution\cite{Chiang:2002vq} with 
the parameterized form Eq.\eqref{eqn:Tag} using 14 parameters (see text).}
\end{figure}

In light of the study of
Ref.\cite{Arndt:2009pc} and the necessity of including the full
multichannel unitarity for the purposes of obtaining a global
description of the hadro- and photoproduction data, we have carried
out an exploratory study to determine the efficacy of doing such a
fit within the Chew-Mandelstam parameterization, Eq.\eqref{eqn:Tag}. 
In the present 
study we perform a coupled-channel fit of the modulus $|\EetaS(W)|$ 
and (the real and imaginary parts of) the existing \said\ and
\maid\ $\pi$--photoproduction amplitudes,
$\EpiS$ in the $S_{11}(1535)$ resonance region, compared in 
Fig.\eqref{fig:picf}. The fit was carried
out by taking the factor $[1-\Kbar C]_{\alpha\sigma}^{-1}$ 
in Eq.\eqref{eqn:Tag} as determined in the
the hadronic study of Ref.\cite{Arndt:2006bf} and adjusting the
parameters of $\Kbar_{\sigma\gamma}$ 
(discussed in detail in the subsections below).
The 
phase of the \EetaS\ multipole in this study gives a resonant wave and
encourages us to continue with this approach, as discussed in the final
section. 

\begin{figure}[t]
\includegraphics[width=250pt,keepaspectratio,clip]{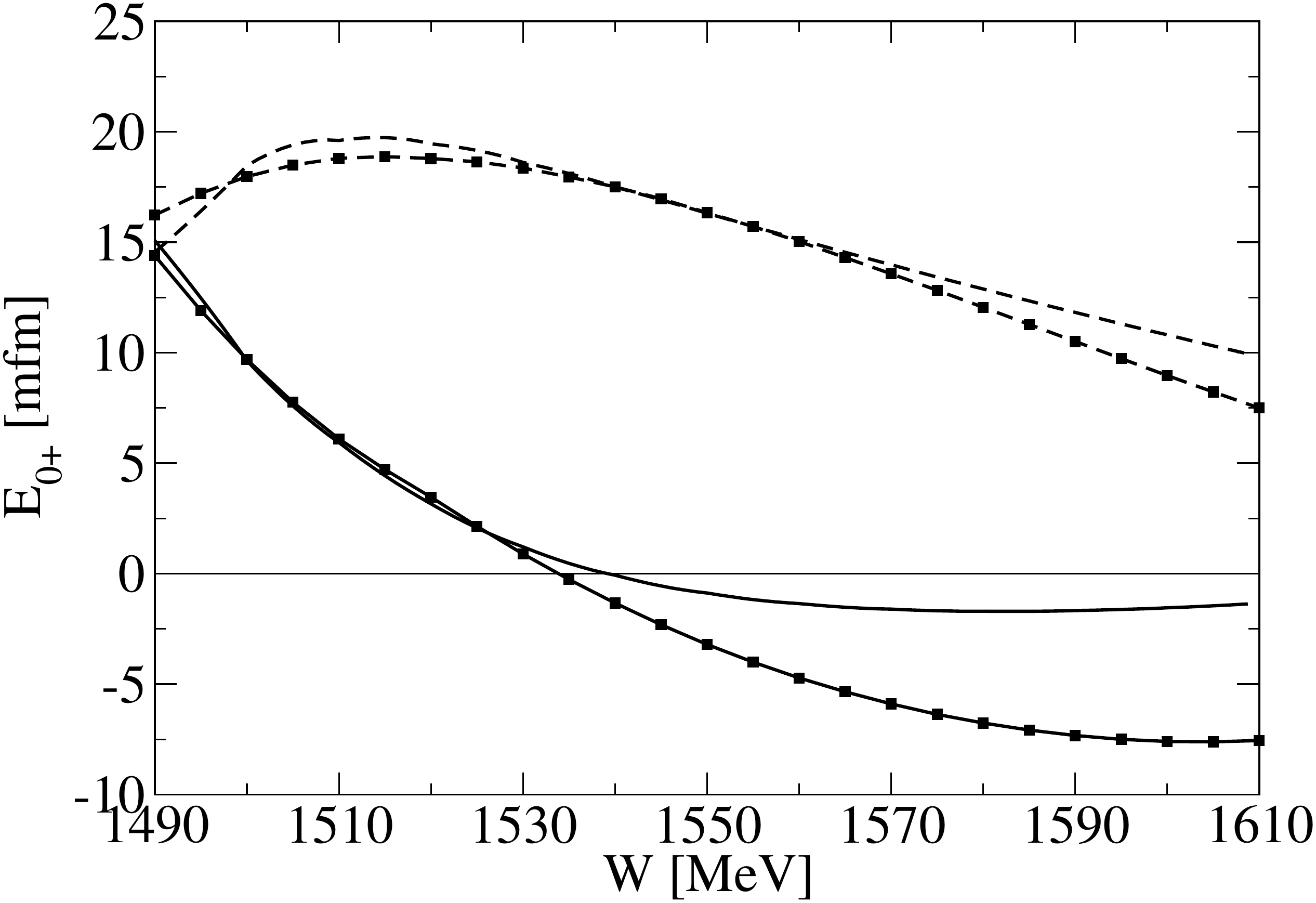}
\caption{\label{fig:E0+MAIDeta-7}The real (solid curves) and imaginary 
(dashed curve) for \EetaS from the 7-parameter fit in Fig.\eqref{fig:E0+MAID-7}
compared with the $\eta$--\maid\ solution\cite{Chiang:2001as}, marked
by squares.}
\end{figure}

The decision to fit the modulus $|\EetaS|$ is based on empirical
considerations. The \maid\cite{Chiang:2001as} parameterization by 
the Dubna-Mainz-Taipei Collaboration (DMT) 
and the model calculations in 
Refs.\cite{Tabakin:1988wi,Kaiser:1996js,Green:1999iq,Aznauryan:2003zg}
agree at the few-percent level on the modulus of
the low-energy $\eta$--photoproduction 
amplitude, $|\EetaS(W)|$. This is anticipated on the grounds that, in 
the $S_{11}(1535)$ resonance region, the differential cross section is 
largely angle independent and therefore dominated by the $S$--wave production. 
It also indicates that the production is largely resonant, but we do not make 
this common assumption.

While the modulus $|\EetaS|$ appears to be known at the level of a few
percent, the $\pi$--photoproduction $S_{11}$ amplitude is, surprisingly, 
not very well determined through different
parameterizations. Figure \eqref{fig:picf} shows the
\said\cite{Arndt:2006bf} and \maid\cite{Chiang:2001as} results
for $\EpiS$. Given this discrepancy, we have also carried
out the fit described above with the modulus $|\EetaS|$ and the \maid\ 
parameterization.

\begin{figure}[b]
\includegraphics[width=250pt,keepaspectratio,clip]{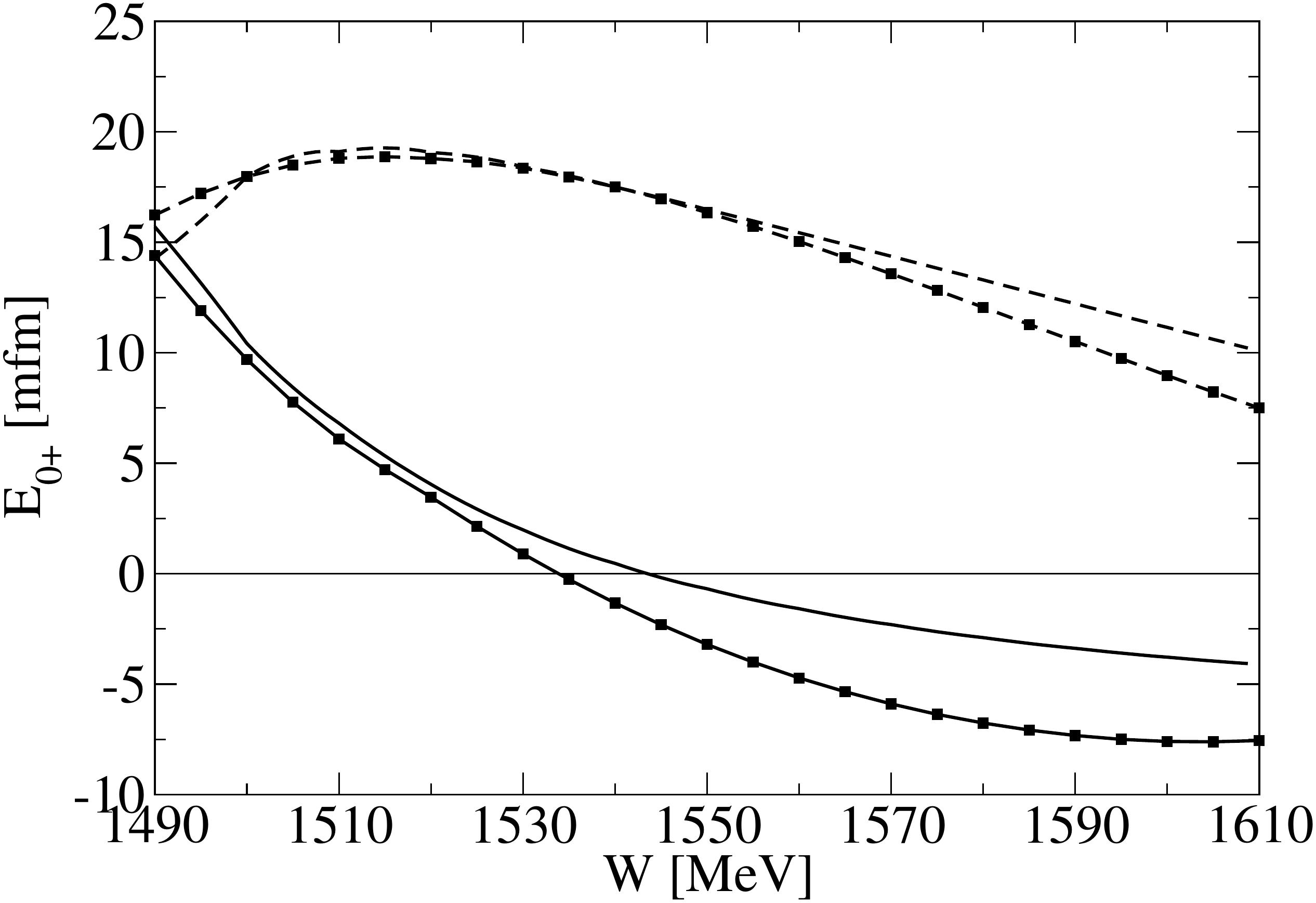}
\caption{\label{fig:E0+MAIDeta-14}The real (solid curves) and imaginary 
(dashed curve) for \EetaS from the 14-parameter fit in 
Fig.\eqref{fig:E0+MAID-14} compared with the $\eta$--\maid\ 
solution\cite{Chiang:2001as}, marked by squares.}
\end{figure}

\subsection{Fit with \said\ \EpiS}
Figure \eqref{fig:Esaid} shows the result of fitting the
modulus $|\EetaS|$ and the real and imaginary parts of the
\said\ \EpiS\ multipole \cite{Arndt:1989ww}
using an eight-parameter fit. The Chew-Mandelstam
\Kbar\ matrix was assumed to have the form
\begin{align}
\Kbar_{\sigma\gamma}(W) &= c_{\sigma\gamma,0} 
                        + c_{\sigma\gamma,1}\zbar_{\sigma\gamma}
\end{align}
taking $n_{\alpha\beta}=1$, in Eq.\eqref{eqn:Kbarpar},
for $\alpha$ and $\beta$ taking values in 
the set of four channels, $\pn,\pD,\rn$, and $\en$. The energy variable
$\zbar_{\alpha\beta}$ is
\begin{align}
\zbar_{\alpha\beta} &= W-W_{t,\alpha},
\end{align}
where the threshold masses, $W_{t,\alpha}$ are $m_\pi+m_N$, $2m_\pi+ m_N$, 
$2m_\pi+ m_N$, and $m_\eta+m_N$ for $\alpha=\pn,\pD,\rn$, and $\en$,
respectively, and $W_{t,\alpha}$ is taken to be the lower of the thresholds
for channels $\alpha$ and $\beta$.
The eight parameters were varied in the fit to a total of 
113 pseudodata points including the modulus $|\EetaS|$
over the energy range 1490 MeV $\le W \le $ 1610 MeV and the amplitude 
$\EpiS$ over the energy range 1120 MeV $\le W \le $1610. The $\chi^2$
per datum over for the fits to the pseudodata, generated with the
\said\ interactive code facility\cite{SAID:ssh}, were less than one
in all of the fits made in this work including those in the region
1120 MeV $\le W \le$ 1490 MeV which are not displayed in order to keep
the figures manageable and focus attention on the $S_{11}(1535)$ resonance
region. The pseudodata were assigned 5\% errors in the fit.

\subsection{Fit with \maid\ \EpiS}
The graphs in Figs.\eqref{fig:E0+MAID-7} and \eqref{fig:E0+MAID-14}
used seven and fourteen parameters, respectively, 
to fit the $|\EetaS|$ and \EpiS amplitudes
from \maid\cite{Chiang:2001as}. The seven-parameter fit in 
Fig.\eqref{fig:E0+MAID-7} is the minimal set of parameters needed
to obtain a $\chi^2$ per datum $\lesssim 1$. The parameters used in
Eq.\eqref{eqn:Kbarpar} for this fit were $c_{\pi\gamma,n}$, $n=0,1,2$,
$c_{\rho\gamma,n}$, where $n=0,1$ and $c_{\eta\gamma,0}$ and $c_{\eta\gamma,1}$.
The quality is degraded at the higher energy end of the
fit region for the imaginary part of \EpiS. Nearly perfect agreement
is obtained if we use a fourteen parameter form for $\Kbar_{\sigma\gamma}$.
The parameters used in
Eq.\eqref{eqn:Kbarpar} for this fit were $c_{\pi\gamma,n}$, 
$c_{\Delta\gamma,n}$, $c_{\rho\gamma,n}$, where $n=0,1,2,3$ 
and $c_{\eta\gamma,0}$ and $c_{\eta\gamma,1}$.

Note, from Figs.\eqref{fig:E0+MAIDeta-7} and \eqref{fig:E0+MAIDeta-14},
that the fit giving the better representation of the \maid\ \EpiS 
amplitude is similarly closer to the \maid\ \EetaS result. This is 
somewhat surprising perhaps, since although the \maid\ pion- and 
eta-photoproduction use the same pole positions in both
amplitudes, these parameterizations are not constrained by unitarity.

\section{Conclusion and ongoing work}
\label{sec:conclusion}
We reviewed the implication of unitarity on the analytic structure of
the single meson production scattering and reaction amplitudes. The
non-analyticities in the regions $W>0$ and $W<0$, the right- and left-hand
cuts, respectively were demonstrated to be properly accounted for by
the $N/D$ approach. We related the Chew-Mandelstam $K$-matrix parameterization
to the $N/D$ approach, showing that the parameterization of the $\Kbar$
matrix neglects the effects of the distant left-hand cut. The purpose of
this review is to place the long-used \said\ amplitudes in the context of
other hadronic amplitude parameterization schemes and to lay the groundwork
for future improvements to the existing parameterization forms.

Using the Chew-Mandelstam $K$ matrix $\Kbar$, we performed a simultaneous
coupled-channel fit of the $\eta$--photoproduction $S_{11}$ multipole 
modulus, $|\EetaS|$ and the $\pi$--photoproduction amplitude, $\EpiS$.
The parameterization was restricted only to the CM $K$ matrix elements
$\Kbar_{\sigma\gamma}$ in Eq.\eqref{eqn:Tag}, while the $[1-\Kbar C]^{-1}$
factors were taken from the existing \said\ fits to the hadronic data. 
The anticipated resonant structure for the phase of the \EetaS\ multipole
was demonstrated in fits to both \said\ and \maid\ amplitudes.

The results of the exploratory study indicate that this is a reasonable
approach toward the objective of
determining a complete set of scattering and reaction amplitudes for
$\pn\to\pn$, $\pn\to\en$, $\gn\to\pn$, and $\gn\to\en$ processes in
a multichannel unitary formalism.
The first stage in this procedure, demonstrating
that coupled-channel simultaneous fits of the $\pi$-- 
and $\eta$--photoproduction
reactions for a single partial wave ($S_{11}$) is possible, 
has been completed.  The next phase, consists of a fit to the
$\pi$-- photoproduction reaction observables.
Following this, a simultaneous fit to 
the reaction observables for the $\pi$-- and $\eta$--photoproduction 
reactions will be performed. As a practical matter, these two phases
will be completed using the $[1-\Kbar C]^{-1}$ ``rescattering'' factors 
determined in separate fits to the hadronic scattering and reaction data.
The final phase of the study will be a simultaneous fit to both the
hadronic and electromagnetic scattering and reaction observables and will
constitute, at least for two-body unitarity, a global description of the
hadro- and photoproduction amplitudes.

\begin{acknowledgments}
The authors thank R.\ Arndt without whom this work would not
be possible. This work was supported in part by the U.S.\ Department 
of Energy Grant DE-FG02-99ER41110. We thank the Department of Energy's 
Institute for Nuclear Theory at the University of Washington for its 
hospitality and the Department of Energy for partial support during the 
initiation of this work.
\end{acknowledgments}

\bibliography{master}

\begin{thebibliography}{63}
\expandafter\ifx\csname natexlab\endcsname\relax\def\natexlab#1{#1}\fi
\expandafter\ifx\csname bibnamefont\endcsname\relax
  \def\bibnamefont#1{#1}\fi
\expandafter\ifx\csname bibfnamefont\endcsname\relax
  \def\bibfnamefont#1{#1}\fi
\expandafter\ifx\csname citenamefont\endcsname\relax
  \def\citenamefont#1{#1}\fi
\expandafter\ifx\csname url\endcsname\relax
  \def\url#1{\texttt{#1}}\fi
\expandafter\ifx\csname urlprefix\endcsname\relax\def\urlprefix{URL }\fi
\providecommand{\bibinfo}[2]{#2}
\providecommand{\eprint}[2][]{\url{#2}}

\bibitem[{\citenamefont{Crede et~al.}(2005)}]{Crede:2003ax}
\bibinfo{author}{\bibfnamefont{V.}~\bibnamefont{Crede}} \bibnamefont{et~al.}
  (\bibinfo{collaboration}{CB-ELSA}), \bibinfo{journal}{Phys. Rev. Lett.}
  \textbf{\bibinfo{volume}{94}}, \bibinfo{pages}{012004}
  (\bibinfo{year}{2005}), \eprint{hep-ex/0311045}.

\bibitem[{\citenamefont{Nakabayashi et~al.}(2006)}]{Nakabayashi:2006ut}
\bibinfo{author}{\bibfnamefont{T.}~\bibnamefont{Nakabayashi}}
  \bibnamefont{et~al.}, \bibinfo{journal}{Phys. Rev.}
  \textbf{\bibinfo{volume}{C74}}, \bibinfo{pages}{035202}
  (\bibinfo{year}{2006}).

\bibitem[{\citenamefont{Williams et~al.}(2009)}]{Williams:2009yj}
\bibinfo{author}{\bibfnamefont{M.}~\bibnamefont{Williams}} \bibnamefont{et~al.}
  (\bibinfo{collaboration}{CLAS}), \bibinfo{journal}{Phys. Rev.}
  \textbf{\bibinfo{volume}{C80}}, \bibinfo{pages}{045213}
  (\bibinfo{year}{2009}), \eprint{0909.0616}.

\bibitem[{\citenamefont{Ajaka et~al.}(1998)}]{Ajaka:1998zi}
\bibinfo{author}{\bibfnamefont{J.}~\bibnamefont{Ajaka}} \bibnamefont{et~al.},
  \bibinfo{journal}{Phys. Rev. Lett.} \textbf{\bibinfo{volume}{81}},
  \bibinfo{pages}{1797} (\bibinfo{year}{1998}).

\bibitem[{\citenamefont{Elsner et~al.}(2007)}]{Elsner:2007hm}
\bibinfo{author}{\bibfnamefont{D.}~\bibnamefont{Elsner}} \bibnamefont{et~al.}
  (\bibinfo{collaboration}{CBELSA}), \bibinfo{journal}{Eur. Phys. J.}
  \textbf{\bibinfo{volume}{A33}}, \bibinfo{pages}{147} (\bibinfo{year}{2007}),
  \eprint{nucl-ex/0702032}.

\bibitem[{\citenamefont{Dugger et~al.}(2004)}]{Dugger:2010etaprop}
\bibinfo{author}{\bibfnamefont{M.}~\bibnamefont{Dugger}} \bibnamefont{et~al.}
  (\bibinfo{collaboration}{CLAS}) (\bibinfo{year}{2004}),
  \urlprefix\url{http://www.jlab.org/exp_prog/proposals/05/PR05-012.ps}.

\bibitem[{\citenamefont{McGeorge et~al.}(2008)}]{McGeorge:2007tg}
\bibinfo{author}{\bibfnamefont{J.~C.} \bibnamefont{McGeorge}}
  \bibnamefont{et~al.}, \bibinfo{journal}{Eur. Phys. J.}
  \textbf{\bibinfo{volume}{A37}}, \bibinfo{pages}{129} (\bibinfo{year}{2008}),
  \eprint{0711.3443}.

\bibitem[{\citenamefont{Cheng et~al.}(1987)\citenamefont{Cheng, Kuo, and
  Li}}]{Cheng:1987yw}
\bibinfo{author}{\bibfnamefont{W.~K.} \bibnamefont{Cheng}},
  \bibinfo{author}{\bibfnamefont{T.~T.~S.} \bibnamefont{Kuo}},
  \bibnamefont{and} \bibinfo{author}{\bibfnamefont{G.~L.} \bibnamefont{Li}},
  \bibinfo{journal}{Phys. Lett.} \textbf{\bibinfo{volume}{B195}},
  \bibinfo{pages}{515} (\bibinfo{year}{1987}).

\bibitem[{\citenamefont{Haider and Liu}(2009)}]{Haider:2009yf}
\bibinfo{author}{\bibfnamefont{Q.}~\bibnamefont{Haider}} \bibnamefont{and}
  \bibinfo{author}{\bibfnamefont{L.~C.} \bibnamefont{Liu}},
  \bibinfo{journal}{Acta Phys. Polon. Supp.} \textbf{\bibinfo{volume}{2}},
  \bibinfo{pages}{121} (\bibinfo{year}{2009}), \eprint{0902.4248}.

\bibitem[{\citenamefont{Prakhov et~al.}(2005)}]{Prakhov:2005qb}
\bibinfo{author}{\bibfnamefont{S.}~\bibnamefont{Prakhov}} \bibnamefont{et~al.},
  \bibinfo{journal}{Phys. Rev.} \textbf{\bibinfo{volume}{C72}},
  \bibinfo{pages}{015203} (\bibinfo{year}{2005}).

\bibitem[{\citenamefont{Basdevant and Berger}(1979)}]{Basdevant:1978tx}
\bibinfo{author}{\bibfnamefont{J.~L.} \bibnamefont{Basdevant}}
  \bibnamefont{and} \bibinfo{author}{\bibfnamefont{E.~L.}
  \bibnamefont{Berger}}, \bibinfo{journal}{Phys. Rev.}
  \textbf{\bibinfo{volume}{D19}}, \bibinfo{pages}{239} (\bibinfo{year}{1979}).

\bibitem[{\citenamefont{Edwards and Thomas}(1980)}]{Edwards:1980sa}
\bibinfo{author}{\bibfnamefont{B.~J.} \bibnamefont{Edwards}} \bibnamefont{and}
  \bibinfo{author}{\bibfnamefont{G.~H.} \bibnamefont{Thomas}},
  \bibinfo{journal}{Phys. Rev.} \textbf{\bibinfo{volume}{D22}},
  \bibinfo{pages}{2772} (\bibinfo{year}{1980}).

\bibitem[{\citenamefont{Arndt et~al.}(1985)\citenamefont{Arndt, Ford, and
  Roper}}]{Arndt:1985vj}
\bibinfo{author}{\bibfnamefont{R.~A.} \bibnamefont{Arndt}},
  \bibinfo{author}{\bibfnamefont{J.~M.} \bibnamefont{Ford}}, \bibnamefont{and}
  \bibinfo{author}{\bibfnamefont{L.~D.} \bibnamefont{Roper}},
  \bibinfo{journal}{Phys. Rev.} \textbf{\bibinfo{volume}{D32}},
  \bibinfo{pages}{1085} (\bibinfo{year}{1985}).

\bibitem[{\citenamefont{Green and Wycech}(1997)}]{Green:1997yia}
\bibinfo{author}{\bibfnamefont{A.~M.} \bibnamefont{Green}} \bibnamefont{and}
  \bibinfo{author}{\bibfnamefont{S.}~\bibnamefont{Wycech}},
  \bibinfo{journal}{Phys. Rev.} \textbf{\bibinfo{volume}{C55}},
  \bibinfo{pages}{2167} (\bibinfo{year}{1997}), \eprint{nucl-th/9703009}.

\bibitem[{\citenamefont{Arndt et~al.}(1998)\citenamefont{Arndt, Green, Workman,
  and Wycech}}]{Arndt:1998nm}
\bibinfo{author}{\bibfnamefont{R.~A.} \bibnamefont{Arndt}},
  \bibinfo{author}{\bibfnamefont{A.~M.} \bibnamefont{Green}},
  \bibinfo{author}{\bibfnamefont{R.~L.} \bibnamefont{Workman}},
  \bibnamefont{and} \bibinfo{author}{\bibfnamefont{S.}~\bibnamefont{Wycech}},
  \bibinfo{journal}{Phys. Rev.} \textbf{\bibinfo{volume}{C58}},
  \bibinfo{pages}{3636} (\bibinfo{year}{1998}), \eprint{nucl-th/9807009}.

\bibitem[{\citenamefont{Green and Wycech}(1999)}]{Green:1999iq}
\bibinfo{author}{\bibfnamefont{A.~M.} \bibnamefont{Green}} \bibnamefont{and}
  \bibinfo{author}{\bibfnamefont{S.}~\bibnamefont{Wycech}},
  \bibinfo{journal}{Phys. Rev.} \textbf{\bibinfo{volume}{C60}},
  \bibinfo{pages}{035208} (\bibinfo{year}{1999}), \eprint{nucl-th/9905011}.

\bibitem[{\citenamefont{Crede et~al.}(2009)}]{Crede:2009zz}
\bibinfo{author}{\bibfnamefont{V.}~\bibnamefont{Crede}} \bibnamefont{et~al.}
  (\bibinfo{collaboration}{CBELSA/TAPS}), \bibinfo{journal}{Phys. Rev.}
  \textbf{\bibinfo{volume}{C80}}, \bibinfo{pages}{055202}
  (\bibinfo{year}{2009}), \eprint{0909.1248}.

\bibitem[{Gre()}]{Green:note}
\bibinfo{note}{{The $K$ matrix approach of Refs.\cite{Green:1997yia,
  Arndt:1998nm, Green:1999iq} include contributions to the hadronic $T$ matrix
  which represent the $\pi\pi\!N$ channel, though the three-body cut is not
  accounted for properly.}}

\bibitem[{\citenamefont{Kaiser et~al.}(1997)\citenamefont{Kaiser, Waas, and
  Weise}}]{Kaiser:1996js}
\bibinfo{author}{\bibfnamefont{N.}~\bibnamefont{Kaiser}},
  \bibinfo{author}{\bibfnamefont{T.}~\bibnamefont{Waas}}, \bibnamefont{and}
  \bibinfo{author}{\bibfnamefont{W.}~\bibnamefont{Weise}},
  \bibinfo{journal}{Nucl. Phys.} \textbf{\bibinfo{volume}{A612}},
  \bibinfo{pages}{297} (\bibinfo{year}{1997}), \eprint{hep-ph/9607459}.

\bibitem[{\citenamefont{Tiator et~al.}(1998)\citenamefont{Tiator, Knochlein,
  and Bennhold}}]{Tiator:1998qp}
\bibinfo{author}{\bibfnamefont{L.}~\bibnamefont{Tiator}},
  \bibinfo{author}{\bibfnamefont{G.}~\bibnamefont{Knochlein}},
  \bibnamefont{and} \bibinfo{author}{\bibfnamefont{C.}~\bibnamefont{Bennhold}},
  \bibinfo{journal}{PiN Newslett.} \textbf{\bibinfo{volume}{14}},
  \bibinfo{pages}{70} (\bibinfo{year}{1998}), \eprint{nucl-th/9802064}.

\bibitem[{\citenamefont{Aznauryan}(2003)}]{Aznauryan:2003zg}
\bibinfo{author}{\bibfnamefont{I.~G.} \bibnamefont{Aznauryan}},
  \bibinfo{journal}{Phys. Rev.} \textbf{\bibinfo{volume}{C68}},
  \bibinfo{pages}{065204} (\bibinfo{year}{2003}), \eprint{nucl-th/0306079}.

\bibitem[{\citenamefont{Arndt et~al.}(2006)\citenamefont{Arndt, Briscoe,
  Strakovsky, and Workman}}]{Arndt:2006bf}
\bibinfo{author}{\bibfnamefont{R.~A.} \bibnamefont{Arndt}},
  \bibinfo{author}{\bibfnamefont{W.~J.} \bibnamefont{Briscoe}},
  \bibinfo{author}{\bibfnamefont{I.~I.} \bibnamefont{Strakovsky}},
  \bibnamefont{and} \bibinfo{author}{\bibfnamefont{R.~L.}
  \bibnamefont{Workman}}, \bibinfo{journal}{Phys. Rev.}
  \textbf{\bibinfo{volume}{C74}}, \bibinfo{pages}{045205}
  (\bibinfo{year}{2006}), \eprint{nucl-th/0605082}.

\bibitem[{\citenamefont{Chiang et~al.}(2002)\citenamefont{Chiang, Yang, Tiator,
  and Drechsel}}]{Chiang:2001as}
\bibinfo{author}{\bibfnamefont{W.-T.} \bibnamefont{Chiang}},
  \bibinfo{author}{\bibfnamefont{S.-N.} \bibnamefont{Yang}},
  \bibinfo{author}{\bibfnamefont{L.}~\bibnamefont{Tiator}}, \bibnamefont{and}
  \bibinfo{author}{\bibfnamefont{D.}~\bibnamefont{Drechsel}},
  \bibinfo{journal}{Nucl. Phys.} \textbf{\bibinfo{volume}{A700}},
  \bibinfo{pages}{429} (\bibinfo{year}{2002}), \eprint{nucl-th/0110034}.

\bibitem[{\citenamefont{Zimmerman}(1961)}]{Zimmerman:1961aa}
\bibinfo{author}{\bibfnamefont{W.}~\bibnamefont{Zimmerman}},
  \bibinfo{journal}{Nuovo Cim.} \textbf{\bibinfo{volume}{21}},
  \bibinfo{pages}{249} (\bibinfo{year}{1961}).

\bibitem[{\citenamefont{Eden}(1952)}]{Eden:1952aa}
\bibinfo{author}{\bibfnamefont{R.~J.} \bibnamefont{Eden}},
  \bibinfo{journal}{Proc. Royal Soc. of London} \textbf{\bibinfo{volume}{210}},
  \bibinfo{pages}{388} (\bibinfo{year}{1952}).

\bibitem[{\citenamefont{Polkinghorne}(1962)}]{Polkinghorne:1962aa}
\bibinfo{author}{\bibfnamefont{J.~C.} \bibnamefont{Polkinghorne}},
  \bibinfo{journal}{Nuovo Cim.} \textbf{\bibinfo{volume}{23}},
  \bibinfo{pages}{360} (\bibinfo{year}{1962}).

\bibitem[{\citenamefont{Boyling}(1964)}]{Boyling:1964aa}
\bibinfo{author}{\bibfnamefont{J.~B.} \bibnamefont{Boyling}},
  \bibinfo{journal}{Nuovo Cim.} \textbf{\bibinfo{volume}{33}},
  \bibinfo{pages}{1356} (\bibinfo{year}{1964}).

\bibitem[{\citenamefont{Bjorken}(1960)}]{Bjorken:1960zz}
\bibinfo{author}{\bibfnamefont{J.~D.} \bibnamefont{Bjorken}},
  \bibinfo{journal}{Phys. Rev. Lett.} \textbf{\bibinfo{volume}{4}},
  \bibinfo{pages}{473} (\bibinfo{year}{1960}).

\bibitem[{\citenamefont{Heitler}(1941)}]{Heitler:1941a}
\bibinfo{author}{\bibfnamefont{W.}~\bibnamefont{Heitler}},
  \bibinfo{journal}{Math.\ Proc.\ Camb.\ Phil.\ Soc.}
  \textbf{\bibinfo{volume}{37}}, \bibinfo{pages}{291} (\bibinfo{year}{1941}).

\bibitem[{\citenamefont{Goldberger and Watson}(1964)}]{Goldberger:1964aa}
\bibinfo{author}{\bibfnamefont{M.~L.} \bibnamefont{Goldberger}}
  \bibnamefont{and} \bibinfo{author}{\bibfnamefont{K.~M.}
  \bibnamefont{Watson}}, \emph{\bibinfo{title}{Collision Theory}}
  (\bibinfo{publisher}{John Wiley and Sons, Inc., New York},
  \bibinfo{year}{1964}).

\bibitem[{\citenamefont{Workman et~al.}(2009)\citenamefont{Workman, Arndt, and
  Paris}}]{Workman:2008iv}
\bibinfo{author}{\bibfnamefont{R.~L.} \bibnamefont{Workman}},
  \bibinfo{author}{\bibfnamefont{R.~A.} \bibnamefont{Arndt}}, \bibnamefont{and}
  \bibinfo{author}{\bibfnamefont{M.~W.} \bibnamefont{Paris}},
  \bibinfo{journal}{Phys. Rev.} \textbf{\bibinfo{volume}{C79}},
  \bibinfo{pages}{038201} (\bibinfo{year}{2009}), \eprint{0808.2176}.

\bibitem[{\citenamefont{Eden et~al.}(1966)\citenamefont{Eden, Landshoff, Olive,
  and Polkinghorne}}]{Eden:1966sm}
\bibinfo{author}{\bibfnamefont{R.~J.} \bibnamefont{Eden}},
  \bibinfo{author}{\bibfnamefont{P.~V.} \bibnamefont{Landshoff}},
  \bibinfo{author}{\bibfnamefont{D.~I.} \bibnamefont{Olive}}, \bibnamefont{and}
  \bibinfo{author}{\bibfnamefont{J.~C.} \bibnamefont{Polkinghorne}},
  \emph{\bibinfo{title}{The Analytic $S$-matrix}}
  (\bibinfo{publisher}{Cambridge University Press, Cambridge},
  \bibinfo{year}{1966}), \bibinfo{note}{pp.\ 231--232}.

\bibitem[{\citenamefont{Weinberg}(1964)}]{Weinberg:1964zz}
\bibinfo{author}{\bibfnamefont{S.}~\bibnamefont{Weinberg}},
  \bibinfo{journal}{Phys. Rev.} \textbf{\bibinfo{volume}{133}},
  \bibinfo{pages}{B232} (\bibinfo{year}{1964}).

\bibitem[{\citenamefont{Mandelstam}(1958)}]{Mandelstam:1958xc}
\bibinfo{author}{\bibfnamefont{S.}~\bibnamefont{Mandelstam}},
  \bibinfo{journal}{Phys. Rev.} \textbf{\bibinfo{volume}{112}},
  \bibinfo{pages}{1344} (\bibinfo{year}{1958}).

\bibitem[{\citenamefont{Frautschi and Walecka}(1960)}]{Frautschi:1960aa}
\bibinfo{author}{\bibfnamefont{S.~C.} \bibnamefont{Frautschi}}
  \bibnamefont{and} \bibinfo{author}{\bibfnamefont{J.~D.}
  \bibnamefont{Walecka}}, \bibinfo{journal}{Phys. Rev.}
  \textbf{\bibinfo{volume}{120}}, \bibinfo{pages}{1486} (\bibinfo{year}{1960}).

\bibitem[{\citenamefont{Frazer and Fulco}(1960)}]{Frazer:1960zz}
\bibinfo{author}{\bibfnamefont{W.~R.} \bibnamefont{Frazer}} \bibnamefont{and}
  \bibinfo{author}{\bibfnamefont{J.~R.} \bibnamefont{Fulco}},
  \bibinfo{journal}{Phys. Rev.} \textbf{\bibinfo{volume}{119}},
  \bibinfo{pages}{1420} (\bibinfo{year}{1960}).

\bibitem[{\citenamefont{Chew and Mandelstam}(1960)}]{Chew:1960iv}
\bibinfo{author}{\bibfnamefont{G.~F.} \bibnamefont{Chew}} \bibnamefont{and}
  \bibinfo{author}{\bibfnamefont{S.}~\bibnamefont{Mandelstam}},
  \bibinfo{journal}{Phys. Rev.} \textbf{\bibinfo{volume}{119}},
  \bibinfo{pages}{467} (\bibinfo{year}{1960}).

\bibitem[{\citenamefont{Queen and Violini}(1974)}]{Queen:1974dr}
\bibinfo{author}{\bibfnamefont{N.~M.} \bibnamefont{Queen}} \bibnamefont{and}
  \bibinfo{author}{\bibfnamefont{G.}~\bibnamefont{Violini}},
  \emph{\bibinfo{title}{Dispersion theory in high-energy physics}}
  (\bibinfo{publisher}{John Wiley \& Sons, Inc., New York},
  \bibinfo{year}{1974}), \bibinfo{note}{pp.\ 13--14}.

\bibitem[{\citenamefont{Arndt}(1968)}]{Arndt:1968sm}
\bibinfo{author}{\bibfnamefont{R.~A.} \bibnamefont{Arndt}},
  \bibinfo{journal}{Phys. Rev.} \textbf{\bibinfo{volume}{165}},
  \bibinfo{pages}{1834} (\bibinfo{year}{1968}).

\bibitem[{\citenamefont{Barut}(1967)}]{Barut:1967ao}
\bibinfo{author}{\bibfnamefont{A.~O.} \bibnamefont{Barut}},
  \emph{\bibinfo{title}{The theory of the scattering matrix}}
  (\bibinfo{publisher}{The Macmillan Company, New York}, \bibinfo{year}{1967}),
  \bibinfo{note}{pp.\ 211--222}.

\bibitem[{\citenamefont{Bjorken and Nauenberg}(1961)}]{Bjorken:1961nd}
\bibinfo{author}{\bibfnamefont{J.~D.} \bibnamefont{Bjorken}} \bibnamefont{and}
  \bibinfo{author}{\bibfnamefont{M.}~\bibnamefont{Nauenberg}},
  \bibinfo{journal}{Phys. Rev.} \textbf{\bibinfo{volume}{121}},
  \bibinfo{pages}{1250} (\bibinfo{year}{1961}).

\bibitem[{\citenamefont{Martin and Spearman}(1970)}]{Martin:1970bb}
\bibinfo{author}{\bibfnamefont{A.~D.} \bibnamefont{Martin}} \bibnamefont{and}
  \bibinfo{author}{\bibfnamefont{T.~D.} \bibnamefont{Spearman}},
  \emph{\bibinfo{title}{Elementary particle theory}} (\bibinfo{publisher}{North
  Holland Publishing Co., Amsterdam}, \bibinfo{year}{1970}).

\bibitem[{\citenamefont{Davidson and Mukhopadhyay}(1990)}]{Davidson:1990yk}
\bibinfo{author}{\bibfnamefont{R.~M.} \bibnamefont{Davidson}} \bibnamefont{and}
  \bibinfo{author}{\bibfnamefont{N.~C.} \bibnamefont{Mukhopadhyay}},
  \bibinfo{journal}{Phys. Rev.} \textbf{\bibinfo{volume}{D42}},
  \bibinfo{pages}{20} (\bibinfo{year}{1990}).

\bibitem[{\citenamefont{Ceci et~al.}(2008)\citenamefont{Ceci, Svarc, Zauner,
  Manley, and Capstick}}]{Ceci:2006jj}
\bibinfo{author}{\bibfnamefont{S.}~\bibnamefont{Ceci}},
  \bibinfo{author}{\bibfnamefont{A.}~\bibnamefont{Svarc}},
  \bibinfo{author}{\bibfnamefont{B.}~\bibnamefont{Zauner}},
  \bibinfo{author}{\bibfnamefont{M.}~\bibnamefont{Manley}}, \bibnamefont{and}
  \bibinfo{author}{\bibfnamefont{S.}~\bibnamefont{Capstick}},
  \bibinfo{journal}{Phys. Lett.} \textbf{\bibinfo{volume}{B659}},
  \bibinfo{pages}{228} (\bibinfo{year}{2008}), \eprint{hep-ph/0611094}.

\bibitem[{Adl()}]{Adler:1964zero}
\bibinfo{note}{{We note that the current parameterization neglects the
  constraints of chiral symmetry near threshold and in the unphysical region.
  Therefore the Adler zero\cite{Adler:1964um} is, too, neglected.}}

\bibitem[{\citenamefont{Babelon et~al.}(1976)\citenamefont{Babelon, Basdevant,
  Caillerie, and Mennessier}}]{Babelon:1976kv}
\bibinfo{author}{\bibfnamefont{O.}~\bibnamefont{Babelon}},
  \bibinfo{author}{\bibfnamefont{J.~L.} \bibnamefont{Basdevant}},
  \bibinfo{author}{\bibfnamefont{D.}~\bibnamefont{Caillerie}},
  \bibnamefont{and}
  \bibinfo{author}{\bibfnamefont{G.}~\bibnamefont{Mennessier}},
  \bibinfo{journal}{Nucl. Phys.} \textbf{\bibinfo{volume}{B113}},
  \bibinfo{pages}{445} (\bibinfo{year}{1976}).

\bibitem[{\citenamefont{Arndt et~al.}(1990{\natexlab{a}})\citenamefont{Arndt,
  Workman, Li, and Roper}}]{Arndt:1989ww}
\bibinfo{author}{\bibfnamefont{R.~A.} \bibnamefont{Arndt}},
  \bibinfo{author}{\bibfnamefont{R.~L.} \bibnamefont{Workman}},
  \bibinfo{author}{\bibfnamefont{Z.}~\bibnamefont{Li}}, \bibnamefont{and}
  \bibinfo{author}{\bibfnamefont{L.~D.} \bibnamefont{Roper}},
  \bibinfo{journal}{Phys. Rev.} \textbf{\bibinfo{volume}{C42}},
  \bibinfo{pages}{1853} (\bibinfo{year}{1990}{\natexlab{a}}).

\bibitem[{\citenamefont{Arndt et~al.}(1990{\natexlab{b}})\citenamefont{Arndt,
  Workman, Li, and Roper}}]{Arndt:1990ej}
\bibinfo{author}{\bibfnamefont{R.~A.} \bibnamefont{Arndt}},
  \bibinfo{author}{\bibfnamefont{R.~L.} \bibnamefont{Workman}},
  \bibinfo{author}{\bibfnamefont{Z.}~\bibnamefont{Li}}, \bibnamefont{and}
  \bibinfo{author}{\bibfnamefont{L.~D.} \bibnamefont{Roper}},
  \bibinfo{journal}{Phys. Rev.} \textbf{\bibinfo{volume}{C42}},
  \bibinfo{pages}{1864} (\bibinfo{year}{1990}{\natexlab{b}}).

\bibitem[{\citenamefont{Arndt et~al.}(2002)\citenamefont{Arndt, Strakovsky, and
  Workman}}]{Arndt:2001si}
\bibinfo{author}{\bibfnamefont{R.~A.} \bibnamefont{Arndt}},
  \bibinfo{author}{\bibfnamefont{I.~I.} \bibnamefont{Strakovsky}},
  \bibnamefont{and} \bibinfo{author}{\bibfnamefont{R.~L.}
  \bibnamefont{Workman}}, \bibinfo{journal}{PiN Newslett.}
  \textbf{\bibinfo{volume}{16}}, \bibinfo{pages}{150} (\bibinfo{year}{2002}),
  \eprint{nucl-th/0110001}.

\bibitem[{\citenamefont{Arndt et~al.}(2004)\citenamefont{Arndt, Briscoe,
  Strakovsky, Workman, and Pavan}}]{Arndt:2003if}
\bibinfo{author}{\bibfnamefont{R.~A.} \bibnamefont{Arndt}},
  \bibinfo{author}{\bibfnamefont{W.~J.} \bibnamefont{Briscoe}},
  \bibinfo{author}{\bibfnamefont{I.~I.} \bibnamefont{Strakovsky}},
  \bibinfo{author}{\bibfnamefont{R.~L.} \bibnamefont{Workman}},
  \bibnamefont{and} \bibinfo{author}{\bibfnamefont{M.~M.} \bibnamefont{Pavan}},
  \bibinfo{journal}{Phys. Rev.} \textbf{\bibinfo{volume}{C69}},
  \bibinfo{pages}{035213} (\bibinfo{year}{2004}), \eprint{nucl-th/0311089}.

\bibitem[{\citenamefont{Koch}(1985)}]{Koch:1985bp}
\bibinfo{author}{\bibfnamefont{R.}~\bibnamefont{Koch}}, \bibinfo{journal}{Z.
  Phys.} \textbf{\bibinfo{volume}{C29}}, \bibinfo{pages}{597}
  (\bibinfo{year}{1985}).

\bibitem[{\citenamefont{Julia-Diaz et~al.}(2007)\citenamefont{Julia-Diaz, Lee,
  Matsuyama, and Sato}}]{JuliaDiaz:2007kz}
\bibinfo{author}{\bibfnamefont{B.}~\bibnamefont{Julia-Diaz}},
  \bibinfo{author}{\bibfnamefont{T.~S.~H.} \bibnamefont{Lee}},
  \bibinfo{author}{\bibfnamefont{A.}~\bibnamefont{Matsuyama}},
  \bibnamefont{and} \bibinfo{author}{\bibfnamefont{T.}~\bibnamefont{Sato}},
  \bibinfo{journal}{Phys. Rev.} \textbf{\bibinfo{volume}{C76}},
  \bibinfo{pages}{065201} (\bibinfo{year}{2007}), \eprint{arXiv:0704.1615
  [nucl-th]}.

\bibitem[{\citenamefont{Shklyar et~al.}(2005)\citenamefont{Shklyar, Lenske,
  Mosel, and Penner}}]{Shklyar:2004ba}
\bibinfo{author}{\bibfnamefont{V.}~\bibnamefont{Shklyar}},
  \bibinfo{author}{\bibfnamefont{H.}~\bibnamefont{Lenske}},
  \bibinfo{author}{\bibfnamefont{U.}~\bibnamefont{Mosel}}, \bibnamefont{and}
  \bibinfo{author}{\bibfnamefont{G.}~\bibnamefont{Penner}},
  \bibinfo{journal}{Phys. Rev.} \textbf{\bibinfo{volume}{C71}},
  \bibinfo{pages}{055206} (\bibinfo{year}{2005}), \eprint{nucl-th/0412029}.

\bibitem[{Tab()}]{Tablenote}
\bibinfo{note}{{The $\chi^2$ per datum analysis presented in Table
  \eqref{tab:chi2} includes results for groups that have provided amplitudes in
  a manner allowing the reconstruction of observables at arbitary angle and
  energy.}}

\bibitem[{\citenamefont{Drechsel et~al.}(2007)\citenamefont{Drechsel, Kamalov,
  and Tiator}}]{Drechsel:2007if}
\bibinfo{author}{\bibfnamefont{D.}~\bibnamefont{Drechsel}},
  \bibinfo{author}{\bibfnamefont{S.~S.} \bibnamefont{Kamalov}},
  \bibnamefont{and} \bibinfo{author}{\bibfnamefont{L.}~\bibnamefont{Tiator}},
  \bibinfo{journal}{Eur. Phys. J.} \textbf{\bibinfo{volume}{A34}},
  \bibinfo{pages}{69} (\bibinfo{year}{2007}), \eprint{0710.0306}.

\bibitem[{\citenamefont{Arndt et~al.}(1995)\citenamefont{Arndt, Strakovsky,
  Workman, and Pavan}}]{Arndt:1995bj}
\bibinfo{author}{\bibfnamefont{R.~A.} \bibnamefont{Arndt}},
  \bibinfo{author}{\bibfnamefont{I.~I.} \bibnamefont{Strakovsky}},
  \bibinfo{author}{\bibfnamefont{R.~L.} \bibnamefont{Workman}},
  \bibnamefont{and} \bibinfo{author}{\bibfnamefont{M.~M.} \bibnamefont{Pavan}},
  \bibinfo{journal}{Phys. Rev.} \textbf{\bibinfo{volume}{C52}},
  \bibinfo{pages}{2120} (\bibinfo{year}{1995}), \eprint{nucl-th/9505040}.

\bibitem[{\citenamefont{Watson}(1952)}]{Watson:1952ji}
\bibinfo{author}{\bibfnamefont{K.~M.} \bibnamefont{Watson}},
  \bibinfo{journal}{Phys. Rev.} \textbf{\bibinfo{volume}{88}},
  \bibinfo{pages}{1163} (\bibinfo{year}{1952}).

\bibitem[{\citenamefont{Workman}(2006)}]{Workman:2005eu}
\bibinfo{author}{\bibfnamefont{R.~L.} \bibnamefont{Workman}},
  \bibinfo{journal}{Phys. Rev.} \textbf{\bibinfo{volume}{C74}},
  \bibinfo{pages}{055207} (\bibinfo{year}{2006}), \eprint{nucl-th/0510025}.

\bibitem[{\citenamefont{Arndt}(2009)}]{Arndt:2009pc}
\bibinfo{author}{\bibfnamefont{R.~A.} \bibnamefont{Arndt}},
  \bibinfo{howpublished}{private communication} (\bibinfo{year}{2009}).

\bibitem[{\citenamefont{Chiang et~al.}(2003)\citenamefont{Chiang, Yang, Tiator,
  Vanderhaeghen, and Drechsel}}]{Chiang:2002vq}
\bibinfo{author}{\bibfnamefont{W.-T.} \bibnamefont{Chiang}},
  \bibinfo{author}{\bibfnamefont{S.~N.} \bibnamefont{Yang}},
  \bibinfo{author}{\bibfnamefont{L.}~\bibnamefont{Tiator}},
  \bibinfo{author}{\bibfnamefont{M.}~\bibnamefont{Vanderhaeghen}},
  \bibnamefont{and} \bibinfo{author}{\bibfnamefont{D.}~\bibnamefont{Drechsel}},
  \bibinfo{journal}{Phys. Rev.} \textbf{\bibinfo{volume}{C68}},
  \bibinfo{pages}{045202} (\bibinfo{year}{2003}), \eprint{nucl-th/0212106}.

\bibitem[{\citenamefont{Tabakin et~al.}(1988)\citenamefont{Tabakin, Dytman, and
  Rosenthal}}]{Tabakin:1988wi}
\bibinfo{author}{\bibfnamefont{F.}~\bibnamefont{Tabakin}},
  \bibinfo{author}{\bibfnamefont{S.~A.} \bibnamefont{Dytman}},
  \bibnamefont{and} \bibinfo{author}{\bibfnamefont{A.~S.}
  \bibnamefont{Rosenthal}} (\bibinfo{year}{1988}), \bibinfo{note}{presented at
  Topical Conf. on Excited Baryons Troy, N.Y., Aug 4-6}.

\bibitem[{SAI()}]{SAID:ssh}
\bibinfo{note}{{The {\sc said} interactive suite of codes is available by
  secure shell interface via: {\tt ssh -C -X said@said.phys.gwu.edu}.}}

\bibitem[{\citenamefont{Adler}(1965)}]{Adler:1964um}
\bibinfo{author}{\bibfnamefont{S.~L.} \bibnamefont{Adler}},
  \bibinfo{journal}{Phys. Rev.} \textbf{\bibinfo{volume}{137}},
  \bibinfo{pages}{B1022} (\bibinfo{year}{1965}).

\end{thebibliography}

\end{document}